\documentclass[manuscript,screen]{acmart}

\AtBeginDocument{
  }

\usepackage{multirow}
\usepackage{listings}
\usepackage{makecell}
\usepackage{booktabs}
\usepackage{tcolorbox}
\usepackage{threeparttable}
\usepackage{algorithm}
\usepackage{algpseudocode}  
\usepackage{amsmath}  
\usepackage{enumitem}
\usepackage{subfigure}
\newtheorem{myDef}{Definition}[section]

\renewcommand\footnotetextcopyrightpermission[1]{}

\lstdefinestyle{JavaStyle}{
    language=Java,
    basicstyle=\footnotesize\ttfamily,
    keywordstyle=\color{blue},
    commentstyle=\color{green!70!black},
    stringstyle=\color{orange},
    numbers=left,
    numberstyle=\tiny\color{gray},
    stepnumber=1,
    numbersep=5pt,
    backgroundcolor=\color{gray!5},
    frame=lines,
    rulecolor=\color{black},
    breaklines=true,
    showstringspaces=false,
    tabsize=2,
    captionpos=b
    escapeinside={(*@}{@*)}, 
}

\author{Taiming Wang}
\affiliation{
  \institution{School of Computer Science \& Technology, Beijing Institute of Technology}
  \city{Beijing}
  \country{China}
  \postcode{100081}}
\email{wangtaiming@bit.edu.cn}
\author{Hui Liu}
\authornote{Corresponding author}
\affiliation{
  \institution{School of Computer Science \& Technology, Beijing Institute of Technology}
  \city{Beijing}
  \country{China}
  \postcode{100081}}
\email{liuhui08@bit.edu.cn}
\author{Yuxia Zhang}
\affiliation{
  \institution{School of Computer Science \& Technology, Beijing Institute of Technology}
  \city{Beijing}
  \country{China}
  \postcode{100081}}
\email{yuxiazh@bit.edu.cn}

\author{Yanjie Jiang}
\authornotemark[1]
\affiliation{
\institution{Key Lab of HCST (PKU), MOE; SCS, Peking University}
\city{Beijing}
\country{China}
}
\email{yanjiejiang@pku.edu.cn}


\begin{document}
\title[Recommending Variable Names for Extract Local Variable Refactorings]{Recommending Variable Names for Extract Local Variable Refactorings}

\begin{abstract}
\emph{Extract local variable} is one of the most popular refactorings. It is frequently employed to replace occurrences of a complex expression with simple accesses to a newly introduced variable that is initialized by the original complex expression. Consequently, most IDEs and refactoring tools provide automated support for this refactoring, e.g., to suggest names for the newly extracted variables. However, we find approximately 70\% of the names recommended by these IDEs are different from what developers manually constructed, adding additional renaming burdens to developers and providing limited assistance. 
In this paper, we introduce \emph{VarNamer}, an automated approach designed to recommend variable names for \emph{extract local variable} refactorings. Through a large-scale empirical study, we identify key contexts, such as variable initializations and homogeneous variables (variables whose initializations are identical to that of the newly extracted variable), that are useful for composing variable names. Leveraging these insights, we developed a set of heuristic rules through program static analysis techniques, e.g., lexical analysis, syntax analysis, control flow analysis, and data flow analysis, and employ data mining techniques, i.e., FP-growth algorithm, to recommend variable names effectively.
Notably, some of our heuristic rules have been successfully integrated into \emph{Eclipse}, where they are now distributed with the latest releases of the IDE. Evaluation of \emph{VarNamer} on a dataset of 27,158 real-world \emph{extract local variable} refactorings in Java applications demonstrates its superiority over state-of-the-art IDEs. Specifically, \emph{VarNamer} significantly increases the chance of exact match by 52.6\% compared to \emph{Eclipse} and 40.7\% compared to \emph{IntelliJ IDEA}. We also evaluated the proposed approach with real-world extract local variable refactorings conducted in C++ projects, and the results suggest that the approach can achieve comparable performance on programming languages besides Java. It may suggest the generalizability of \emph{VarNamer}. Finally, we designed and conducted a user study to investigate the impact of \emph{VarNamer} on developers’ productivity. The results of the user study suggest that our approach can speed up the refactoring by 27.8\% and reduce 49.3\% edits on the recommended variable names.  
\end{abstract}


\maketitle
\section{Introduction}\label{sec:Introduction}
\emph{Extract local variable} is a well-known and widely used refactoring~\cite{MurphyHill,Negara2013}.  It involves replacing one or more occurrences of a complex expression with a newly added variable and direct access to the variable.  This refactoring simplifies the involved source code and enhances readability and maintainability by providing meaningful variable names for extracted expressions.  Consequently, \emph{extract local variable} is frequently employed by developers~\cite{golubev2021one}, with studies reporting that it accounts for over 80\% of refactorings conducted with automated tool support~\cite{Negara2013}. This underscores its significance in software development for improving code maintainability and comprehensibility.

Existing IDEs (e.g., \emph{Eclipse}~\cite{Eclipse}, IntelliJ \emph{IDEA}~\cite{IDEA}, NetBeans~\cite{NetBeans}, and Visual Studio~\cite{VisualStudio}) offer automated or semi-automated support for \emph{extract local variable} refactoring, typically comprising three main components. The first component validates whether the selected expression can be extracted as a new local variable and determines which occurrences of the expression can be replaced with variable accesses~\cite{Chi2023}. The second component is to recommend a name for the newly introduced variable, and the third component executes the refactoring by automatically modifying the code based on decisions made in the preceding steps.
While IDEs are often accurate in precondition validation and source code modification, variable name recommendation, the second component, tends to be less accurate, with reported accuracy rates of less than 30\% in our evaluation. 
One reason for the inaccuracy of variable name recommendations in IDEs is the emphasis on efficiency without substantial latency. IDEs prioritize fast heuristic-based approaches for variable name recommendation, often at the expense of accuracy. For instance, \emph{Eclipse} employs a series of heuristic rules based on three types of contexts: the initialization itself, the parameters assigned by the initialization, and the data type of the initialization.
Another contributing factor is the oversight of crucial contexts by existing IDEs. Our empirical study revealed that IDEs often overlook critical context such as homogeneous variables, which are variables with initializations identical to the expressions being extracted. This lack of context-sensitivity contributes to the low accuracy in variable name recommendations, creating a gap between recommended names and those expected by developers.
Inaccurate recommendations not only burden developers but also lead to low-quality variable names that can diminish the readability and maintainability of source code~\cite{butler2009relating,butler2010exploring}. Therefore, there is a pressing need to improve the accuracy of variable name recommendations in IDEs to enhance developer productivity and maintain code quality.

To address the limitations of existing IDEs in accurately recommending variable names for the extract local variable refactorings, we introduce an automated approach called \emph{VarNamer}. \emph{VarNamer} leverages the surrounding contexts of the refactoring, particularly focusing on the initialization of the variable and the names of its homogeneous variables. Our approach begins with an empirical study aimed at identifying the most informative contexts for constructing variable names. By analyzing a large corpus of code, we determine that the initialization of the variable and the names of its homogeneous variables are particularly valuable in this regard. Based on these findings, \emph{VarNamer} comprises three components: reuse-based name recommendation, generation-based name recommendation, and name selection. The name reuse component leverages the presence of homogeneous variables to suggest suitable names, while the name generation component utilizes the initialization context to generate relevant name candidates. Finally, the name selection component selects the most appropriate name from the generated candidates.
To evaluate the performance of \emph{VarNamer}, we constructed a dataset of 27,158 real-world \emph{extract local variable} refactorings mined from the commit histories of 1,000 GitHub repositories. Our evaluation results demonstrate that \emph{VarNamer} significantly outperforms the state-of-the-art IDEs, achieving a 52.6\% improvement in exact name matching compared to \emph{Eclipse} and a 40.7\% improvement compared to \emph{IntelliJ IDEA}.
To assess the extensibility of \emph{VarNamer}, we manually collected a dataset comprising 50 real-world extract local variable refactorings from prominent C++ open-source projects. We then evaluated the C++ version of \emph{VarNamer} on this dataset, and the results indicate that our approach achieves comparable performance across different programming languages. Furthermore, we conducted user experiments where developers utilized our approach to perform the extract local variable refactorings. The results reveal a notable improvement in both time efficiency and edit efficiency. Specifically, our approach facilitated a 27.8\% reduction in refactoring time and a 49.3\% decrease in the number of edits required for recommended variable names. These findings underscore the practical utility of \emph{VarNamer} in enhancing the developers' productivity and satisfaction with the recommended variable names.
Notably, the implementations of the key heuristic rules proposed in this paper have been successfully merged into the mainstream IDE \textit{Eclipse} and are now distributed with its latest releases.  
The contributions of this paper are as follows:
\begin{itemize}
    \item We conducted a large-scale empirical study on variable name recommendation for the extract local variable refactorings, marking the first of its kind in this research domain.
    \item We introduced \emph{VarNamer}, a heuristics-based approach designed to recommend variable names for the extract local variable refactorings. Notably, the key heuristic rules proposed in \emph{VarNamer} have been integrated into mainstream IDE \textit{Eclipse}.
    \item 
    We curated two datasets comprising real-world extract local variable refactorings, one containing 27,158 instances from Java programs and the other containing 50 instances from C++ programs. These datasets, along with the replication package, have been made publicly available~\cite{VarNamer}.
\end{itemize} 

The rest of this paper is structured as follows. Section~\ref{sec:Background} presents the definitions of key terminologies in this paper with examples. Section~\ref{sec:EmpiricalStudy} presents how we conducted the empirical study and the corresponding results.   Section~\ref{sec:Approach} presents the design details of \emph{VarNamer}, and the evaluation results are presented in Section~\ref{sec:Evaluation}. Section~\ref{sec:Discussion} discusses the threats to validity and the limitations of this paper. We review the related work in Section~\ref{sec:RelatedWork}, and finally Section~\ref{sec:Conclusion} concludes this paper.

\section{Background}\label{sec:Background}
This section provides an essential foundation for readers to establish a common understanding of the concepts discussed throughout the paper. In this section, we introduce five key terminologies by explaining concepts and giving definitions alongside illustrative examples.

\begin{figure}[]
    \centering
    \includegraphics[width=\linewidth,clip]{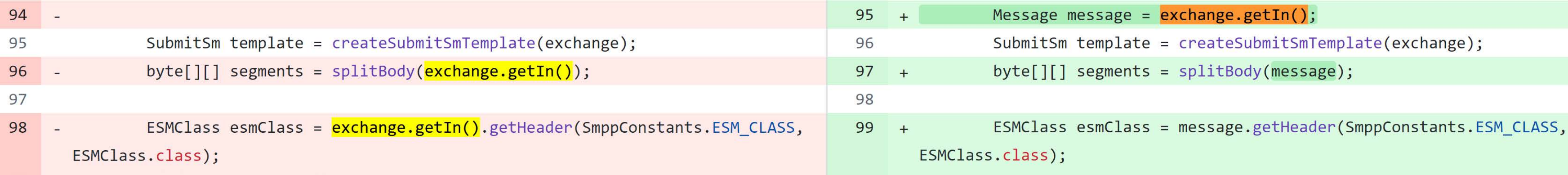}
    \caption{An Example of Extract Local Variable Refactoring}
    \label{fig:ExtractLocalVariable}
    \vspace{-1em}
\end{figure}

\emph{\textbf{Extract local variable refactoring}} (also known as \emph{introduce local variable refactoring}~\cite{ExtractVariable}) is a widely used refactoring technique that replaces expressions with a newly introduced local variable and its references. When expressions become complex and difficult to interpret, it is better to extract them as variables with meaningful names to enhance their self-explanatory nature. After extraction, multiple occurrences of the expression can be replaced with references to the new variable. Another benefit of this refactoring is the removal of duplicate expressions that appear multiple times within a single method, thereby reducing the complexity of the enclosing method and ensuring the expression is executed only once. This benefit is particularly desirable when the repeated expression is lengthy or resource-consuming (e.g., invoking a computation-intensive method).
A typical example of extract local variable refactoring is shown in Fig.~\ref{fig:ExtractLocalVariable}. Here, the expression \emph{"exchange.getIn()"}~\cite{RefactoringExample1} (highlighted in yellow) appears twice within a single method (lines 96 and 98 on the left). Conducting this refactoring, replacing the expression with a local variable named "message" in lines 97 and 99 on the right, reduces the complexity of the enclosing method and improves its brevity.

\begin{lstlisting}[float,style=JavaStyle,caption={Example of Homogeneous Variable.},label={code}]
// Method where the extract local variable refactoring happened.
default String packageName() {
-   return name().substring(0, `\textbf{name().lastIndexOf('.')}`); 
+    final int `\textbf{dotIdx}` = `\textbf{name().lastIndexOf('.')}`;
+    if (dotIdx < 0) {
+        return "";
+    } else {
+        return name().substring(0, `\textbf{dotIdx}`);
+    }
}
// Sibling method where the homogeneous variable is retrieved
default String simpleName() {
    `\textbf{// HomogeneousVariable: dotIdx}`
    final int `\textbf{dotIdx}` = `\textbf{name().lastIndexOf('.')}`; 
    if (dotIdx < 0) {
        return name();
    } else {
        return name().substring(dotIdx + 1);
    }
}
\end{lstlisting}

\emph{\textbf{Initialization}} of a local variable refers to the expression used to initialize the variable. During an extract local variable refactoring, developers typically introduce a local variable and initialize it with the extracted expression. 
A typical example is illustrated in Fig.~\ref{fig:ExtractLocalVariable}. In this example, the \emph{initialization} is \emph{"exchange.getIn()"} (highlighted in yellow) in lines 96 and 98. Notably, the \emph{initialization} serves as the most direct context for recommending the variable name. During our empirical study and recommendation process, we only consider the identifier tokens of the initializations, such as \emph{"exchange getIn"}.

\emph{\textbf{Data type}} of an \emph{initialization} refers to the data type of the newly introduced variable after the refactoring. During an extract local variable refactoring, developers sometimes use the data type as the name of the newly introduced variable.
A typical example is \emph{"Message"} in line 95 of Fig.~\ref{fig:ExtractLocalVariable}. Similar to the preprocessing of the \emph{initialization}, we only retain the identifier tokens of the \emph{data type}.

\emph{\textbf{Assignment}} represents the relationship where parameters, variables, or fields are assigned by an \emph{initialization}. According to an empirical study conducted by Liu et al.~\cite{7886980}, there can be high lexical similarity between method arguments and parameters. In such cases, the names of these assigned objects can be considered as part of the variable names.
For example, in line 96 of Fig.~\ref{fig:ExtractLocalVariable}, the \emph{initialization} \emph{"exchange.getIn()"} is assigned as an argument to the method \emph{"splitBody"} whose complete signature is \emph{"protected byte[][] splitBody(Message message)"}. Here, the \emph{assignment} corresponds to the formal parameter name of the method \emph{"splitBody"}, which is "message".

\emph{\textbf{Declaration Context}} represents the \textbf{data type} and \textbf{initialization} within the variable declaration statement (i.e., except for the variable name itself). This is the most intuitive context for variable name recommendation. Thus most of the popular IDEs, e.g., Eclipse and IDEA, have developed heuristic rules to recommend variable names according to such declaration contexts.

\begin{myDef}
\textbf{Homogeneous Variable:} A variable that shares the same initialization expression as the \emph{initialization} intended for extraction. These \emph{homogeneous variables} are crucial contexts for recommending names during the extract local variable refactoring process. However, they are often overlooked by the current implementations of popular IDEs, e.g., \emph{Eclipse} and \emph{IDEA}. \emph{Homogeneous variables} can exist within various scopes, including the same Java file where the refactoring occurs, in Java files within the same package, or in Java files within the same project.

\end{myDef}
A typical example of \emph{homogeneous variable} is presented in Listing~\ref{code}. The first method in this example, named \emph{"packageName"}, is where the refactoring occurred, while the second method, named \emph{"simpleName"}, is a sibling method in the same Java file as \emph{"packageName"}. The \emph{Initialization}, highlighted in bold font (lines 3 and 4), is \emph{"name().lastIndexOf('.')"}.
The ground truth name, provided by the original developer, is \emph{"dotIdx"} (line 4). Interestingly, at the time of the refactoring, a variable named \emph{"dotIdx"} (line 14) already exists in \emph{"simpleName"} and is initialized by the \emph{Initialization}. Since developers have already assigned a name to the to-be-extracted \emph{Initialization}, we can reuse this name for the recommendation. This variable, due to its high similarity with the newly extracted variable, is referred to as a \emph{homogeneous variable}.

\section{Empirical Study}\label{sec:EmpiricalStudy}
In this section, we present the methodologies and findings of the empirical study conducted to explore the most influential factors in variable name recommendation for the extract local variable refactorings. Through this study, we aim to identify and analyze the contexts that contribute most significantly to the effectiveness of variable name recommendations. It is worth noting that we do not discriminate \emph{token} and \emph{sub-token} in this paper, and they both refer to a word in identifiers, e.g., \emph{"simple"}  in \emph{"simpleName"}.

\subsection{Research Questions}
The empirical study should answer the following  questions:
\begin{itemize}
    \item \textbf{RQ1:} How often could the name tokens of the extracted variables be found in the contexts of the variables?
    
    \begin{itemize}
        \item \textbf{RQ1-1:} What are the possible contexts of the variables where the tokens of variable names can be found?
        \item \textbf{RQ1-2:} How often could the name tokens be found in different types of contexts?
    \end{itemize}
    
    \item \textbf{RQ2:} How accurate is it to recommend variable names by simply copying (reusing) the names of homogeneous variables?
\end{itemize}

RQ1 concerns where we can retrieve the tokens to compose the complete name for the newly extracted variable, and how often we can find them in different types of contexts. We further split RQ1 into two sub-questions, i.e., RQ1-1 and RQ1-2. To address RQ1-1, we conducted a sample analysis to identify
the most influential factors (i.e., context) contributing to variable name recommendation. The investigation results suggest that there are four types of essential contexts. Consequently, we extended our analysis to a larger dataset and examined how often can the name tokens be found in these four types of context (RQ1-2). The investigation results suggest that homogeneous variables are critical and they often contain the desired name tokens. Inspired by such a finding, we investigated  RQ2 with a detailed analysis of the challenges associated with copying names from homogeneous variables, providing insights for the development of heuristic rules in leveraging this context effectively. Answering RQ1 and RQ2 would significantly facilitate the design of context-based approaches to recommending variable names for the extract local variable refactorings.

\subsection{Methodology}
\subsubsection{Data Collection}\label{subsubsec:DataCollection}
\begin{figure}[]
    \centering
    \includegraphics[width=\linewidth, clip]{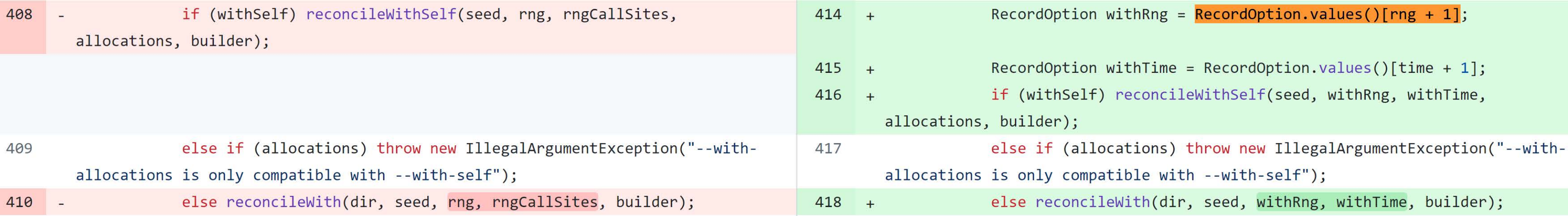}
    \caption{An Example of A False Positive}
    \label{fig:FP}
    \vspace{-1em}
\end{figure}

To conduct the empirical study, we first constructed a dataset by collecting real-world extract local variable refactorings from open-source projects on GitHub~\cite{GitHub}. We selected the top 1,000 open-source Java projects on GitHub sorted by stars. This selection criteria of projects was made to ensure a diverse and representative sample of real-world refactorings. From each of the selected projects, we collected the extract local variable refactoring as follows:
\begin{itemize}
\item First, we leveraged RefactoringMiner~\cite{RefactoringMiner} to discover a list of \textit{potential} extract local variable refactorings (noted as $pRs$) that have been conducted in the given project. We selected RefactoringMiner because it represents the state-of-the-art method in the automated discovery of refactoring histories~\cite{RefactoringMiner2}. 
\item Second, we filtered out false positives in $pRs$ automatically with static analysis. A \textit{potential} extract variable refactoring was deemed a false positive if the extracted expression did not appear in the original version of the source code (i.e., before refactoring) or if it was not replaced with the newly introduced variable in the new version (i.e., after refactoring). An example is presented in Fig.~\ref{fig:FP}. The \emph{initialization} (highlighted in orange) was \emph{"RecordOption.values()[rng+1]"}~\cite{FalsePositiveExample}.  RefactoringMiner reported it as an extract local variable refactoring because a new local variable \emph{"withRng"} was created, and it was used in the original expression (line 416 on the right and line 408 on the left). However, we noticed that this expression did not exist before the refactoring was conducted (i.e., line 408 on the left does not contain this expression). The modification was more like an addition of new variables (lines 414 and 415 on the right) and bug fixing with the new variables (lines 416 and 418 on the right) than an extract local variable refactoring. 
\item Finally, we removed duplicate refactorings caused by branch merge. If a refactoring was conducted on one commit that was located in branch A, the refactoring would be recorded by another commit where branch A was merged with the main branch. As a result, the same refactoring would be discovered twice, resulting in duplicate refactorings. 
\end{itemize}

We finally obtained 32,039 real-world extract local variable refactorings from 745 projects. Such refactorings were further divided into two disjointed datasets: $EmpiricalDataSet$ and $TestingDataSet$. The former was composed of 4,881 extract local variable refactorings discovered from randomly selected 100 projects. It was used for the empirical study in this section. $TestingDataSet$ was composed of the other refactorings (27,158 refactorings in 645 projects) and was employed for evaluation in Section~\ref{sec:Evaluation}.  

\subsubsection{Investigation of RQ1}\label{subsubsec:RQ1Methodology}
To answer RQ1-1, we randomly sampled 50 extract local variable refactorings from $EmpiricalDataSet$, and  analyzed each of them as follows:
\begin{itemize}
    \item We leveraged spacy~\cite{spaCy} to split the name of the extracted variable into sub-tokens.
    \item For the sub-tokens in the variable name, we located all their occurrences within the Java file where the refactoring happened. 
    \item For each occurrence of the sub-token, two authors independently identified the type of the context (e.g., variable initialization, the data type of the variable, and homogeneous variables) that contained the sub-token. The participants were Ph.D. students majoring in computer science, with more than 4 years of Java programming experience and more than 3 years of research experience in software analysis. The Cohen's kappa coefficient~\cite{cohen1960coefficient} is 0.87, indicating a strong agreement between the two raters. Any discrepancies were resolved through discussion until a consensus was reached. 
\end{itemize}

With the sample analysis introduced in the preceding paragraph, we can identify the most critical contexts for variable name recommendation. As indicated by the results in Section~\ref{subsec:RQ1}, these essential contexts include the initialization of the extracted variable (referred to as \emph{initialization}), \emph{homogeneous variables}, the \emph{data type} of the extracted variable, and \emph{assignments}. Understanding and leveraging these contexts are fundamental for developing effective strategies for recommending variable names in extract local variable refactorings.
To investigate RQ1-2, we further validated the findings inferred from the small sample set (50 refactorings) and examined the frequency of name tokens of variables appearing in the four types of contexts across the entire $EmpiricalDataSet$. Note that $EmpiricalDataSet$ comprised 4,881 refactorings.  The validation process proceeded as follows:
\begin{itemize}
    \item First, we obtained variable names and their corresponding contexts, including \textit{initialization}, \textit{homogeneous variables}, \textit{data type}, and \textit{assignments}, through static code analysis using \emph{Eclipse JDT} ~\cite{JDT}. This resulted in a list of 5-tuples: [Variable Name, Initialization, Homogeneous Variable, Data Type, Assignment]. Note that we parsed the resulting initialization and data type to remove the operators and separators, e.g., "\textit{.}", "\textit{(}", and "\textit{)}" leveraging javalang~\cite{javalang} because such tokens would not appear in variable names.
    \item Second, we used spacy~\cite{spaCy} to split the items in the 5-tuple into sub-tokens. Subsequently, all sub-tokens were converted to lowercase, consistent with previous name recommendation studies~\cite{Liu2019, Nguyen2020, Li2021}.
    \item Third, by comparing the sub-tokens in a given variable name against sub-tokens in its contexts, for each type of context we obtained a list whose size is the number of the sub-tokens in the variable name. For the example presented in listing~\ref{code}, the variable name (i.e., \emph{"dotIdx"}) had 2 sub-tokens (\emph{"dot"} and \emph{"idx"}). The corresponding result list is [[0,0],[1,1],[0,0],[0,0]], where \emph{"1"} represented that the sub-token could be found in the corresponding context, and \emph{"0"} otherwise. It is worth noting that for a single refactoring, there is only one \emph{initializations} and one \emph{data type}. However, it may have multiple  \textit{homogeneous variables} and multiple \emph{assignments}. In these cases, we concatenated all names in  \textit{homogeneous variables} (or  \emph{assignments}) with a blank space, e.g., \emph{\{dot idx dot idx dot idx\}}. 
    
    \item Fourth, we further investigated how often the to-be-recommended variable names were identical to the contexts. Unlike the above token analysis, we did not perform the splitting, and thus each unit is a variable name instead of one sub-token. We employ a methodology similar to the above analysis, which we omit here to avoid repetition.
\end{itemize}

\subsubsection{Investigation of RQ2}\label{subsubsec:RQ2Methodology}

To answer RQ2, we conducted the empirical study as follows. First, we investigated how the distance between the newly extracted variable and its homogeneous variables influenced the performance of recommendations. We categorized distances into three levels: within the enclosing project, within the enclosing package, and within the enclosing document, i.e., the Java file. Subsequently, we quantitatively assessed the impact of distance-based filtering on the performance of recommendations.
Second, we observed that not all retrieved homogeneous variable names perfectly matched those of the newly extracted variables. To determine when it is appropriate to reuse the names of homogeneous variables and to explore how we can strategically select the most promising homogeneous variables for variable name recommendation, we randomly selected 100 refactorings from our dataset, comprising 50 successful and 50 failed cases for qualitative analysis. Successful cases refer to instances where the names of homogeneous variables are identical to the variable names. In this case, reusing these names will result in a successful recommendation. Conversely, failed cases occur when the names of homogeneous variables differ from the variable names, indicating that reusing these names will not lead to a successful recommendation. Following a standard brainstorming methodology, we derived four special cases that need to be carefully addressed. To ensure the reliability of these results, we repeated this task on a new dataset. To be specific, we collected 20 additional Java open-source projects that were not included in the original set of 1000 projects. Using RefactoringMiner~\cite{RefactoringMiner}, we identified the extract variable refactorings from their commit history. We then filtered out the invalid refactorings using the same method as in our previous data construction process (Section~\ref{subsubsec:DataCollection}), resulting in 5,374 valid refactorings. From such refactorings, we sampled another 100 refactorings, consisting of 50 successful cases and 50 failed cases, to repeat the task. As a result, we still identified the four special cases illustrated in Section~\ref{subsubsec:CaseStudy}. By refining the study design and analyzing both quantitative and qualitative aspects, we gained insights into the factors influencing the effectiveness of variable name recommendations based on homogeneous variables.

\subsection{RQ1: Useful Contexts for Variable Name Recommendation}\label{subsec:RQ1}
\subsubsection{RQ1-1: Useful Contexts}
Based on the manual analysis of 50 real-world extract local variable refactorings as introduced in Section~\ref{subsubsec:RQ1Methodology}, we have identified four useful contexts (refer to Section~\ref{sec:Background} for details) that in total contain 84 sub-tokens composing the names of the extracted local variables.

With the examples presented in Fig.~\ref{fig:ExtractLocalVariable} and Listing~\ref{code}, we illustrate how we identified possible contexts containing the name tokens of variables. We first examined the most intuitive and direct context: \emph{declaration context}, i.e., \emph{initialization} plus \emph{data type}, for name recommendation. In the two examples above, the sub-tokens of the expected variable names are \emph{\{"message"\}} and \emph{\{"dot", "Idx"\}}. Although neither the \emph{initialization} contains any sub-tokens, the \emph{data type} in the first example is  \emph{"Message"}, which is identical (ignoring case) to the expected variable name. 

Intuitively, the extracted variable should fit into the statement from which it was extracted. Consequently, besides the \emph{declaration context}, we also examined the statement where the \emph{initialization} is extracted.  We found that the expected variable name sometimes matches the parameter name (i.e., \emph{assignments}) if it is assigned by the \emph{initialization}. In the example presented in Fig.~\ref{fig:ExtractLocalVariable}, \emph{"exchange.getIn()"} is assigned as an argument to the method \emph{"splitBody"}, and the corresponding parameter name is \emph{"message"}, which is identical to the expected variable name. 

We then extended our search scope to other parts of the enclosing method and even the entire Java file. We found that sometimes there are variables with the same name initialized by \emph{initialization} elsewhere in the Java file, which we refer to as \emph{homogeneous variables}. An example is shown in  Listing~\ref{code}, where the \emph{homogeneous variable} found in method \emph{"simpleName"} contains all the sub-tokens of the expected variable name, i.e., \emph{\{"dot", "Idx"\}}.

Following the above procedures, we identified all the name tokens in the above four contexts. Specifically, 61.9\% of the name tokens could be found in the \emph{initialization} context, 13.1\% in the names of \emph{homogeneous variables}, 27.4\% in the \emph{data type} context, and 2.4\% in the \emph{assignment} context. It's worth noting that a single sub-token could simultaneously appear in different contexts. 
\begin{tcolorbox}

        \textit{Answer to RQ1-1:}
        The most useful contexts for variable name recommendation include \emph{declaration context} (i.e., \emph{initialization} plus \emph{data type}), \emph{homogeneous variable}, and \emph{assignment}. 
    \end{tcolorbox}

\begin{table*}[]
\renewcommand\arraystretch{1.2}  
    \centering
    \caption{Useful Contexts for Variable Name Recommendation}
    \begin{tabular}{@{}ccccccc@{}}
    \toprule
    & Initialization    & \makecell[c]{Homogeneous \\Variable}   & Data Type    & Assignment &\makecell[c]{Declaration \\Context}  & All Context\\ \midrule
    \# Hitting         & 4,720   & 1,155       & 2,229   & 81     & 5,146  & 5,592 \\
    Hitting Rate(\%)       & 57.3  & 14.0    & 27.1  & 1.0  & 62.5 & 68.0 \\\bottomrule
    \end{tabular}
    \label{tab:RatioAnalysis}
    \end{table*}
    
\begin{table}[]
 \renewcommand\arraystretch{1.2}  
    \centering
    \caption{Chance of Exact Match}
    \renewcommand\arraystretch{1.2}  
    \begin{tabular}{@{}ccccc@{}}
    \toprule
          & Initialization & \makecell[c]{Homogeneous Variable}    & Data Type & Assignment \\ \midrule
    \#Exact Match    & 17                 &  603             & 674      & 21          \\
    Chance of Exact Match (\%)  & 0.3              & 12.4           & 13.8   & 0.4          \\ \bottomrule
    \end{tabular}
    \label{tab:TotalSame}
    \vspace{-1em}
    \end{table}
    
\subsubsection{RQ1-2: Frequency of Name Tokens in Different Context}
We then extended our analysis across the $EmpiricalDataSet$, and the results are presented in Table~\ref{tab:RatioAnalysis}. The first line denotes different types of context. The second line denotes the hitting number, i.e., the number of sub-tokens that can be found in this context. The last line denotes the hitting rate which is calculated by the hitting number divided by the total number of the sub-tokens in all the variable names in $EmpiricalDataSet$ (8,231).  From Table~\ref{tab:RatioAnalysis}, we observe that:
\begin{itemize}
  \item We have a great chance (68.0\%) to find sub-tokens for the to-be-named variable from the given four categories of contexts.
  \item The declaration contexts present a significant opportunity, with a 62.5\% likelihood of containing sub-tokens of the variable name. However, by including additional contexts such as \emph{homogeneous variables} and \emph{assignments}, the likelihood can be substantially improved by 8.8\%=(68.0\% - 62.5\%) / 62.5\%.
  \item \emph{Homogeneous variables} and \emph{data types} have considerable chances (14.0\% and 27.1\%) to contain sub-tokens of the variable name.
\end{itemize}
Overall, the above findings underscore the importance of considering additional contexts, such as \emph{homogeneous variables}, in variable name recommendations for the extract local variable refactorings. 

We also examined the frequency of exact matches between the to-be-recommended variable names and their respective contexts. The results are summarized in Table~\ref{tab:TotalSame}. Here, \emph{\#Exact Match} represents the number of refactorings where the variable names perfectly match the corresponding context. The \emph{chances of Exact Match} is calculated by dividing the \emph{\#Exact Match} by the size of $EmpiricalDataSet$ (4,881).
From Table~\ref{tab:TotalSame} we observe that:
 
\begin{itemize}
    \item \emph{Homogeneous variables} and \emph{data types} exhibit the highest likelihoods (12.4\% and 13.8\%, respectively) of being identical to the variable name. This is primarily because \emph{homogeneous variables} and \emph{data types} often share names with the variables themselves, suggesting that recommending variable names by reusing those of \emph{homogeneous variables} and \emph{data types} could be straightforward and accurate. 
    \item Despite over half (57.3\%) of the sub-tokens of variable names being found in the \emph{initializations} (as shown in  Table~\ref{tab:RatioAnalysis}), the rate of variable names identical to \emph{initializations} is much lower (0.3\%),  This suggests the complexity involved in extracting tokens from \emph{initializations}.
    \item \emph{Assignments} exhibit a low rate of exact match (0.4\%), consistent with its performance of sub-token occurrences (1.0\%).
\end{itemize}

In conclusion, our study sheds light on the complexities and opportunities in variable name recommendation for the extract local variable refactorings. While \emph{initializations} often harbor a significant number of sub-tokens, the process of extracting them accurately can pose challenges. On the other hand, \emph{homogeneous variables}, despite containing fewer sub-tokens, offer a straightforward and reliable resource for variable name recommendation. This suggests that leveraging \emph{homogeneous variables} may present a simpler and more accurate approach compared to relying solely on \emph{initializations}. 

\begin{tcolorbox}
        
        \textit{Answer to RQ1-2:}
        The name tokens of variables appear most frequently in \emph{declaration contexts}, with \emph{initializations} being the most common, followed by \emph{data types}. \emph{Homogeneous variables} also play a significant role, while \emph{assignments} contribute the least. However, in terms of the likelihood of exact matches, \emph{homogeneous variables} and \emph{data types} exhibit the highest probabilities. 
        
    \end{tcolorbox}

\subsection{RQ2: Reuse Names of Homogeneous Variables}\label{subsec:RQ2}
\subsubsection{Searching Scope}\label{subsubsec:SearchScope}
In Table~\ref{tab:SearchScope}, we present statistics regarding the search for homogeneous variables across different scopes, as introduced in Section~\ref{subsubsec:RQ2Methodology}. The first row delineates three scopes: the same project, the same package, and the same document, i.e., Java files, from which homogeneous variables can be retrieved. The second row, i.e., "\textit{\# fruitful cases}", indicates the number of cases where at least one homogeneous variable can be found within the specified distance. The third row, i.e., "\textit{\# correct cases}", denotes the number of cases where correct variable names can be selected from all homogeneous variables. The fourth row, i.e., $P_{correct}$, denotes the probability of selecting the correct variable name among all homogeneous variables, i.e., 
\begin{equation}
    P_{correct}=\frac{\# \ correct \  cases}{\# \ fruitful \ cases}
\end{equation}
The fifth row illustrates the average number of retrieved homogeneous variables, while the sixth row depicts the average time cost of searching for homogeneous variables within the given search scope.

From Table~\ref{tab:SearchScope}, we make the following observations: 
\begin{itemize}
    \item Firstly, as the scope narrows down from the enclosing project to the enclosing package/document, the number of fruitful cases decreases from 1,822 to 1,326 and 879, respectively. However, the possibility of selecting the correct variable name among all homogeneous variables improves from 72.2\% to 75.6\% and 79.7\%. This phenomenon occurs because a closer distance between the recommended variable and its homogeneous variables increases the likelihood of them playing similar roles in their context, thereby sharing the same name.
    \item Secondly, in terms of efficiency, narrowing down the scope notably reduces the overwhelming number of homogeneous variables (83,960.6 vs. 1,209.7 and 2.6) that require further exclusion and identification. Consequently, the time cost significantly decreases (2,250.8ms vs. 116.1ms and 0.6ms).
    \item It's worth noting that the involved projects in $EmpiricalDataSet$ are large, with an average of 2,701 Java files per project. Consequently, searching all files within the enclosing project for homogeneous variables can be time-consuming. Conversely, confining the search within a single file can significantly reduce the computational cost.
\end{itemize}

In conclusion, by narrowing down the scope to the same project, package, or Java file, we observe a trade-off between the number of fruitful cases and the efficiency of the selection process. Given our objective of integrating our approach into mainstream IDEs, we have decided to limit the search scope to the same Java file during the evaluation. This choice is motivated by the desire to optimize the efficiency of our approach within the context of typical development workflows. By focusing on the same Java file, we aim to minimize computational overhead while still providing meaningful and accurate variable name recommendations directly within the developer's immediate coding environment.

\begin{table}[]
\renewcommand\arraystretch{1.2}  
    \centering
    \caption{Homogeneous Variables in Different Scopes}
    \renewcommand\arraystretch{1.2}
    
    \begin{tabular}{@{}lccc@{}}
    \toprule
        Metrics                            &  Project & Package &  Document\\ \midrule
    \# Fruitful Cases               & 1,822                      & 1,326                  & 879              \\
    \# Correct Cases              & 1,316                     & 1,002                 & 701              \\
    $P_{correct}$ (\%)                       & 72.2                    & 75.6                 & 79.7                      \\ 
    Avg. \# Homogeneous Variable    & 83,960.6                   & 1,209.7                & 2.6                \\
    Avg. Time Cost (ms)                  & 2,250.8                  & 116.1                & 0.6                  \\\bottomrule
    \end{tabular}
    \label{tab:SearchScope}
    \vspace{-1em}
    \end{table}

\subsubsection{Special Cases}\label{subsubsec:CaseStudy}
As outlined in Section~\ref{subsubsec:RQ2Methodology}, we have conducted a qualitative analysis to refine the accuracy of reusing the names of \emph{homogeneous variables}. This analysis has yielded valuable insights into scenarios where it is appropriate or inappropriate to reuse such names, contributing to the enhancement of our variable name recommendation approach. Here are the four typical cases we found:
\begin{description}
    \item[Case1] 
    When an \emph{initialization} is a \emph{universal initialization}, reusing the names of \emph{homogeneous variables} is not advisable. \emph{Universal initializations} are those that have been used to initialize different variables (with different names) and have appeared extensively across multiple projects. For instance, \emph{"null"} serves as a \emph{universal initialization} since it is prevalent in initializing any newly created instances. Similarly, \emph{"0"} and \emph{"1"} are prevalent in initializing any Integer objects. In addition, \emph{"true"} and \emph{"false"} are standard choices to initialize any Boolean objects. Finally, using \emph{"new StringBuilder()"} is standard for creating an instance of \emph{"StringBuilder"} across different projects, although the variable names may vary. In such cases, reusing names from \emph{homogeneous variables} may lead to inaccuracies, as these variables may not necessarily resemble the newly extracted variable.
      
    \item[Case2] If the \emph{initialization} is long enough in character length, it is advisable to reuse their names. Longer initializations are typically more complex and specific in functionality, reducing the likelihood of finding homogeneous variables with similar initializations. Therefore, if any such rare and specific homogeneous variables are found, it is highly reliable to reuse their names.
    
    \item[Case3] When the \emph{homogeneous variable} and the variable to be extracted play similar roles within the same statement-level context, reusing their names is advisable. The statement-level context refers to the parent statement where the variables are referenced. For instance, in the example provided in Listing~\ref{code}, both the to-be-extracted variable and its \emph{homogeneous variable} are referenced within statements such as \emph{"return name().substring(0, dotIdx);"} (line 8) and \emph{"return name().substring(dotIdx + 1);"} (line 18), exhibiting structural and literal similarity. Therefore, reusing the name of such a \emph{homogeneous variable} can achieve high accuracy.
    \item[Case4] 
    When the \emph{homogeneous variable} and the variable to be extracted serve similar purposes within the method-level context, reusing their name is advisable. For instance, in the example provided in Listing~\ref{code}, the bodies of the enclosing methods, namely \emph{"packageName"} and \emph{"simpleName"}, exhibit significant structural and literal similarity. This similarity in the method-level context suggests that the \emph{homogeneous variable} can serve as a reliable reference, and reusing its name may yield high accuracy.
\end{description}

\begin{tcolorbox}
        
        \textit{Answer to RQ2:}
        Simply copying (reusing) the names of homogeneous variables has great potential for recommending correct variable names. The possibility of finding a correct variable name among homogeneous variable names (i.e., $P_{correct}$) increases as the search scope narrows down. Simply copying names from homogeneous variables has a success rate ranging from 72.2\% to 79.7\%.
        
    \end{tcolorbox}

\section{Approach}\label{sec:Approach}
\subsection{Overview}
\begin{figure}[]
    \centering
    \includegraphics[width=0.87\linewidth, trim = {0 40 0 50},clip]{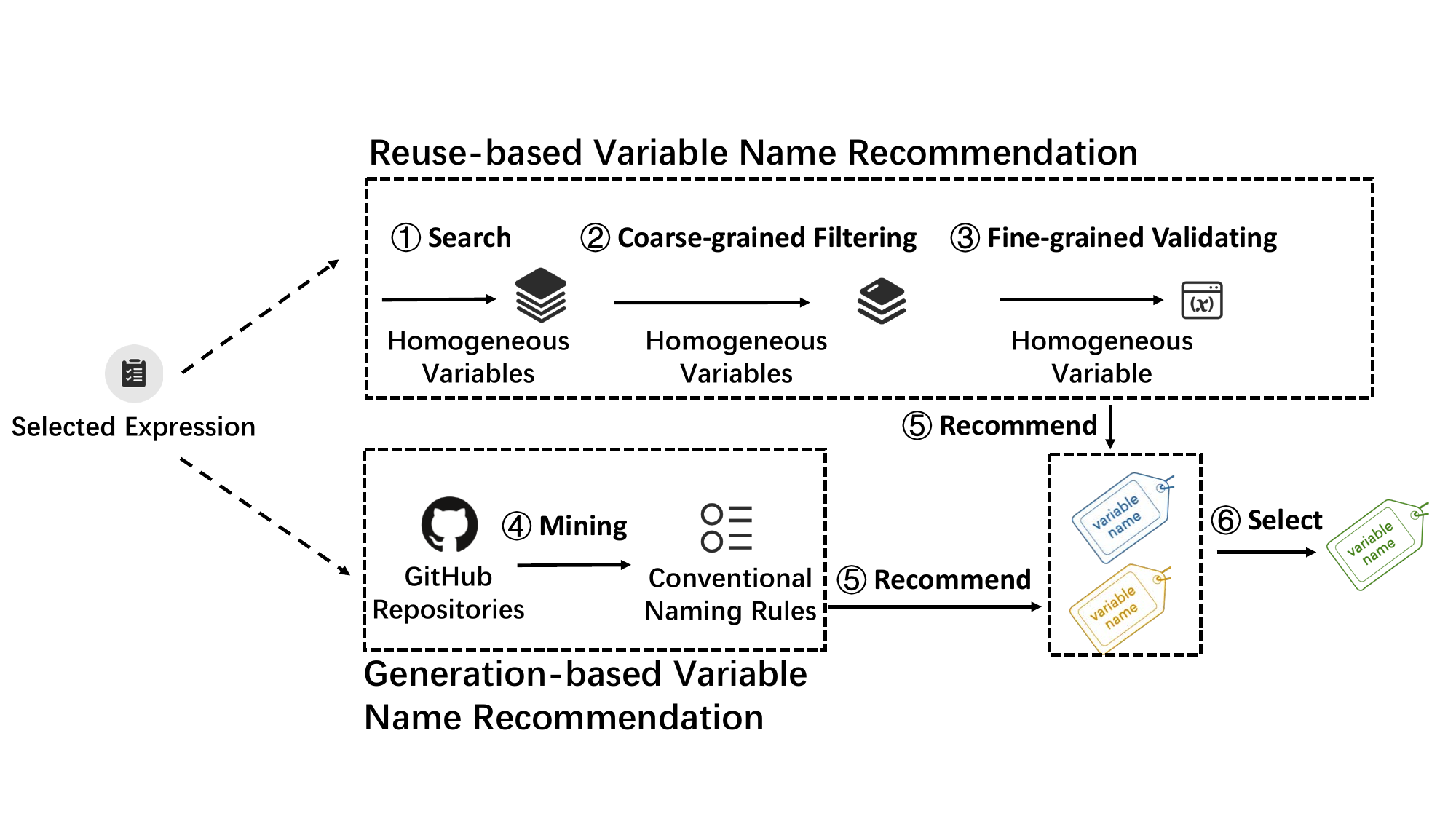}
    \caption{Overview of \textbf{VarNamer}}
    \label{fig:overview}
    \vspace{-1em}
\end{figure}

The overview of \emph{VarNamer} is depicted in Fig.~\ref{fig:overview}. It comprises three key components: rule-based variable name recommendation, generation-based variable name recommendation, and the final name selection and recommendation. For brevity, these components are referred to as name reuse, name generation, and name selection throughout the paper. Initially, \emph{VarNamer} searches for \emph{homogeneous variables} within the enclosing Java file. Upon retrieval, it applies filtering and validation techniques to identify reliable candidates. Simultaneously, \emph{VarNamer} employs conventional naming rules extracted from high-quality code corpora to generate potential names. Finally, a series of heuristic rules are leveraged by \emph{VarNamer} to select the recommended name for developers. Detailed implementation specifics will be discussed in subsequent sections.

\subsection{Reuse-Based Variable Name Recommendation}\label{subsec:NameReuse}
Through our investigation of real-world data, we have observed that blindly reusing the names of homogeneous variables in all cases may not always yield satisfactory results. To maximize the utility of homogeneous variables, we have devised a two-pronged approach consisting of a coarse-grained filter and a fine-grained validator, addressing common scenarios identified in Section~\ref{subsubsec:CaseStudy}.
\subsubsection{Coarse-grained Filter}\label{subsubsec:CoarseFilter}
The coarse-grained filter is tailored to tackle \textbf{Case1} outlined in Section~\ref{subsubsec:CaseStudy}.
We identify \emph{universal initializations} through the following steps:
First, we gather initializations assigned with two or more distinct names across each project in $MiningDataSet$ (refer to Section~\ref{subsubsec:MiningDataSetConstruction}). This results in a set of \emph{project-specific universal initializations}. Subsequently, we deem a \emph{project-specific universal initialization} as a \emph{universal initialization} if it appears extensively across multiple projects. The parameter $ProjectNum$, denoting the number of projects, is configurable (refer to Section~\ref{subsec:ExperimentalSetup}).
The coarse-grained filter operates as follows: it initially verifies whether the initialization qualifies as one of the \emph{universal initializations}. If it does, the name reuse component is deactivated, and the name generation component is utilized instead. Conversely, if the initialization does not meet the criteria for \emph{universal initializations}, it undergoes further validation by the subsequent fine-grained validator to determine whether to reuse the names of \emph{homogeneous variables}.

\subsubsection{Fine-grained Validator}\label{subsubsec:FineFilter}
\begin{figure}[]
    \centering
    \includegraphics[width=0.8\linewidth, trim = {40 70 0 50},clip]{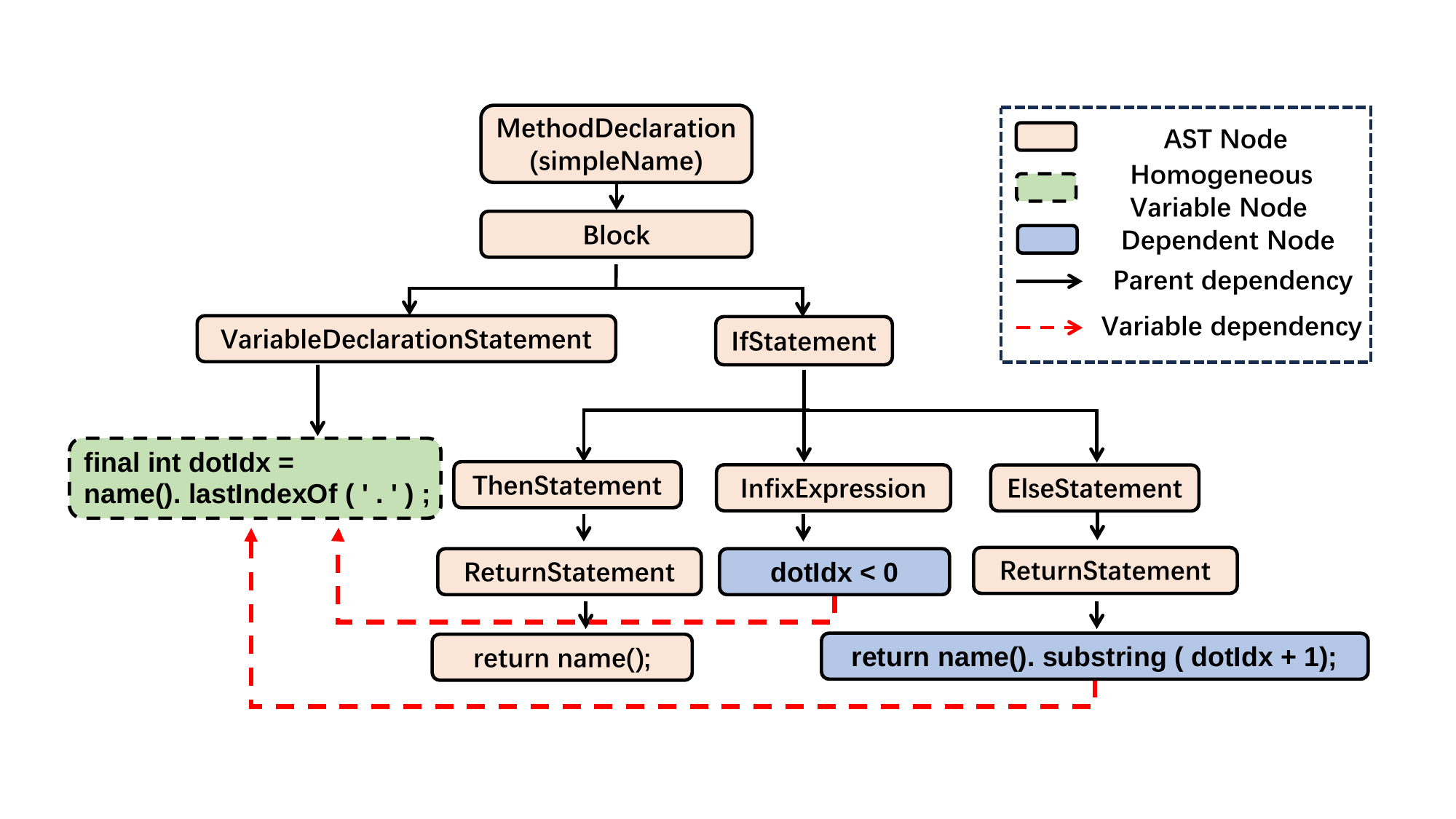}
    \caption{Variable Dependency Graph}
    \label{fig:VDG}
    \vspace{-1em}
\end{figure}

The fine-grained validator is devised to handle \textbf{Case2}, \textbf{Case3}, and \textbf{Case4}, as delineated in Section~\ref{subsubsec:CaseStudy}.
To identify \textbf{Case2}, we compute the character length of the \emph{initialization}s and establish a tunable parameter $IniLength$ (refer to Section~\ref{subsec:ExperimentalSetup}) to distinguish reliable homogeneous variables from unreliable ones. If the character length of the \emph{initialization} exceeds $IniLength$, it is deemed a reliable homogeneous variable; otherwise, further validation is performed to assess if it falls under \textbf{Case3} or \textbf{Case4}.
For \textbf{Case3}, we measure the similarity of the statement-level context (i.e., parent statement where the variables are referenced) between the to-be-extracted variable and the \emph{homogeneous variable}. If the similarity surpasses a tunable parameter $FGSim$, it is considered reliable. Regarding \textbf{Case4}, we assess the similarity between the enclosing method of the homogeneous variable and that of the to-be-extracted variable. If multiple homogeneous variables persist after the aforementioned filtering and validation, only the one with the highest method similarity is retained.
The measurement of these two similarities is outlined as follows. For clarity, we denote the to-be-extracted variable as \emph{ev}, the parent statement where \emph{ev} is extracted as \emph{pev}, and the homogeneous variable as \emph{hv}.
Initially, we created a Variable Dependency Graph (VDG) for each \emph{hv}. A VDG, constructed over the Abstract Syntax Tree (AST) structure, represents the variable dependency relationships, delineating where the variable has been accessed or referenced. An example of VDG is presented in Fig.~\ref{fig:VDG}, and its corresponding code snippet is presented in listing~\ref{code} (lines 12-20).  
A VDG is constructed this way:
\begin{itemize}                           
    \item Initially, we parsed the sibling method where we found \emph{hv}, denoted as the \emph{MethodDeclaration} named \emph{"simpleName"}, to generate an abstract syntax tree leveraging JDT~\cite{JDT}.
    \item Subsequently, we identified the AST node declaring \emph{hv} and the AST nodes dependent on \emph{hv} through static code analysis techniques.
    \item Finally, we augmented the original AST by adding variable dependency edges in addition to the simple parent-child edges, resulting in a VDG for \emph{hv}. As illustrated in Fig.~\ref{fig:VDG}, homogeneous variable (\emph{hv}) named \emph{"dotIdx"} is represented by the green box with a dotted border (line 14 in Listing~\ref{code}). The variable dependent nodes, which are statement-level AST nodes that have a reference (i.e., access) to the \emph{hv}, are depicted with blue boxes (lines 15 and 18 in Listing~\ref{code}). Note that a special case occurs when the variable dependent node appears within a control structure, such as an \emph{if}, \emph{for}, or \emph{while} statement. In these cases, the variable dependent nodes correspond to the conditions of these control structures (i.e., ``dotIdx < 0'' in Fig.~\ref{fig:VDG}). These nodes are connected to the \emph{hv} via variable dependency edges, represented by a red dotted line.  
\end{itemize}

We then explain how the fine-grained validator works with this example:
First, we construct the VDG for \emph{hv} (\emph{"dotIdx"}), which is found in the sibling method \emph{"simpleName"}, and traverse the VDG to obtain the dependent nodes via equation~\ref{eq:traverse}. 
Second, we measure the structural similarity (equation~\ref{eq:structuralSim}) and literal similarity (equation~\ref{eq:literalSim}) of $pev$ and each $vdn$. Third, we measure the context similarity by equation~\ref{eq:totalSim}, and the highest context similarity is selected if there are multiple $vdn$s. If the context similarity is greater than a tunable parameter $FGSim$, we consider $hv$ reliable.
Finally, if there are multiple reliable $hv$s, we calculate the method similarity in the same way (equations~\ref{eq:structuralSim}~\ref{eq:literalSim}~\ref{eq:totalSim}) and the one with the highest method similarity is selected.
Note that the calculation of structural similarity is inspired by Liu et al.~\cite{liu2023automated,fluri2007change,falleri2014fine}. They deemed two AST nodes identical when the AST node type and value were the same.  We loosen the logic and deem two AST nodes as identical if the AST node type is the same.
The literal similarity is measured by normalized Levenshtein distance~\cite{yujian2007normalized}.
Through calculation, the similarities between two $vdn$s and \emph{pev} are 0.24 and 0.67. Finally, we select the latter one (i.e., the highest one) to represent the final context similarity. Due to 0.67 > $FGSim$ (refer Table~\ref{tab:Parameters}), \emph{VarNamer} will reuse the name of the homogeneous variable (i.e., \emph{"dotIdx"}), which is identical to the name given by the original developers. It is worth noting that the reuse strategy has been submitted as pull requests to the \emph{Eclipse} community, and they have been merged into the mainstream~\cite{PR1,PR2}.

\begin{equation}\label{eq:traverse}
    \{\textbf{Variable Dependent Nodes}\}= Traverse(VDG(\texttt{hv}))
\end{equation}
where VDG(\texttt{hv}) denotes the variable dependency graph of \texttt{hv}; Traverse(*) is a function that traverses the given VDG to obtain all the dependent nodes;

\begin{equation}\label{eq:structuralSim}
\setlength{\abovedisplayskip}{2pt}
\setlength{\belowdisplayskip}{2pt}
    \begin{split}
    {\forall\ \textbf{vdn}}  \in \{\textbf{Variable Dependent Nodes}\}, \\
        \textbf{Structural Similarity (\texttt{pev},\texttt{vdn}) }= dice(\texttt{pev},\texttt{vdn}) \\
        = 2 * \frac{\lvert nodes(\texttt{pev}) \cap nodes(\texttt{vdn})\rvert}
        {\lvert nodes(\texttt{pev})\rvert    +   \lvert nodes(\texttt{vdn})\rvert} 
        \rightarrow [0,1]
    \end{split}
\end{equation}
where the $pev$ and $vdn$ are both statement-level AST nodes; $nodes(*)$ is a function to return the sub-tree associated with the AST node $*$, more specifically, all descendant nodes of $*$. nodes(\texttt{pev}) $\cap$ nodes(\texttt{vdn})
denote the common nodes (i.e., share the same node type).
        
\begin{equation}\label{eq:literalSim}
\setlength{\abovedisplayskip}{2pt}
\setlength{\belowdisplayskip}{2pt}
    \begin{split}
    {\forall\ \textbf{vdn}}  \in \{\textbf{Variable Dependent Nodes}\}, \\
        \textbf{Literal Similarity (\texttt{pev},\texttt{vdn})} = 1 - \frac{Levenshtein(\texttt{pev},\texttt{vdn})}{max(length(\texttt{pev}),length(\texttt{vdn}))}\\
        \rightarrow [0,1]
    \end{split}
\end{equation}
where \emph{Levenshtein} is the function to calculate the Levenshtein distance between two texts (i.e., the entire line of the statements without any modifications), and \emph{max} is the function to pick out the larger length of two texts; \emph{length} denotes the character length of the expression.
        
\begin{equation}\label{eq:totalSim}
    \begin{split}
   {\forall\ \textbf{vdn}}  \in \{\textbf{Variable Dependent Nodes}\}, \\
   \textbf{Context Similarity (\texttt{pev},\texttt{vdn})}= \frac{1}{2} * Structural Similarity(\texttt{pev},\texttt{vdn}) \\ 
   + \frac{1}{2} * Literal Similarity(\texttt{pev},\texttt{vdn})\\
   \rightarrow [0,1]
    \end{split}
\end{equation}

\subsection{ Generation-Based Variable Name Recommendation}\label{subsec:NameGeneration}
In this section, we employ data mining techniques to explore the conventions and patterns associated with variable naming, particularly focusing on the relationship between variable names and their corresponding initializations.
Through data mining, we aim to uncover insights into how programmers typically name variables based on how they are initialized. This investigation will provide an understanding of common practices and conventions in variable naming, which facilitates the development of effective naming recommendation tools.

\subsubsection{Data Construction}\label{subsubsec:MiningDataSetConstruction}
We utilized a dataset consisting of 430 open-source projects (GitHub repositories) originally collected by Liu et al.~\cite{Liu2019} for our mining dataset. These projects were sourced from four prominent communities—Apache, Spring, Hibernate, and Google—and were selected based on having a minimum of 100 commits. This criterion ensures that the projects are well-maintained and likely exhibit a higher quality of variable naming practices.
For reference, the list of project names and their corresponding URLs can be accessed on our online repository~\cite{JavaRepos}.
We built our dataset for mining as follows:
\begin{itemize}
    \item To prevent data leakage, we excluded projects utilized in the empirical study (Section~\ref{sec:EmpiricalStudy}) and testing (Section~\ref{sec:Evaluation}) from the 430 projects obtained from Liu et al.~\cite{Liu2019}. This curation resulted in 374 projects, encompassing 326,050 Java files, all of which were utilized for data mining purposes. 
    \item Next, we extracted local variable declarations with initializations from these Java files using JDT~\cite{JDT}. This extraction process yielded a dataset, denoted as $MiningDataSet$, comprising a total of 1,608,843 variable declarations.
    \item Finally, recognizing that the specific features of various types of initializations may influence variable naming conventions, we partitioned $MiningDataSet$ based on the AST node type of the initializations. This segmentation facilitated the subsequent mining procedures.
\end{itemize}

\begin{algorithm}[h]  
    \caption{Mine Conventional Naming Rules.}  
    \label{alg:mining}  
    \begin{algorithmic}[1]  
      \Require  
          $MiningDataSet$: The Dataset for Mining.
       \Ensure 
          $NamingRules$: The Conventional Naming Rules. 
         \State [($NameTokens$,$IniTokens$)] = preprocess($MiningDataSet$)
         \State $Alignments$ = AlignAndReplace([($NameTokens$,$IniTokens$])
         \For {$Alignment \in Alignments$ }
            \If {($Alignment$.contains($placeholder$))}
            \State $ValidAlignments$.add($Alignment$)
            \EndIf
         \EndFor
         \For {$Alignments \in \{ValidAlignments\}$ }
             \State $FPTree$= CreateTree($Alignments$, $MinSupport$)
             \State $CandidateRules$ = GenerateAssociationRules($FPTree$, $MinConfidence$)
             \State $NamingRules$ = Validate($CandidateRules$)
         \EndFor

    \end{algorithmic}  
  \end{algorithm}

\subsubsection{Mining Conventional Naming Rules}\label{subsubsec:MiningNamingPatterns}
In this section, we investigated the conventional rules developers tend to follow when naming variables based on their initializations by employing association analysis techniques. The mining process is outlined in Algorithm~\ref{alg:mining}. Here's a step-by-step breakdown: 
\begin{itemize}
    \item \textbf{Preprocessing:} We preprocessed the variable names and their initializations using the following steps:
    \begin{itemize}
        \item We utilized javalang~\cite{javalang} to extract only identifier tokens from variable names and their initializations.
        \item We used spacy~\cite{spaCy} to split the identifier tokens into sub-tokens, resulting in a 2-tuple [$NameTokens$, $IniTokens$] (line 1).
    \end{itemize}
    \item \textbf{Alignment:} For each 2-tuple, we aligned $NameTokens$ and $IniTokens$ to identify the sub-tokens that appear in both sets. Subsequently, we replaced the aligned sub-tokens in initializations with a placeholder, excluding variable names consisting of only one letter (line 2). For instance, if we had an initialization \emph{"checkConfig.getMessages()"} and its corresponding variable name \emph{"messages"}, the alignment resulted in $alignment$: [\emph{"check"}, \emph{"config"}, \emph{"get"}, $placeholder$].
    \item \textbf{Validation:} We reserved cases where the alignments and replacements succeed, resulting in a collection $ValidAlignments$ (lines 3-7).
    \item \textbf{Mining with FP-growth:} We employed the FP-growth algorithm~\cite{Han2000} (lines 8-12), known for its efficiency and reliability in association rules mining~\cite{Kotsiantis2006}. The algorithm mined frequently co-appearing items (\emph{FCI}) in $ValidAlignments$, yielding $CandidateRules$. It's worth noting that initializations with different AST node types were analyzed separately.
    \item \textbf{Manual Validation:} We manually validated the $CandidateRules$ to ensure their reasonability, resulting in the final $NamingRules$. The parameters $MinSupport$ and $MinConfidence$ were configurable and determined the minimum frequency of items in the \emph{FPTree} and the conditional probability of an \emph{FCI}. These parameters are discussed in detail in Section~\ref{subsec:ExperimentalSetup}.
\end{itemize}
Some examples of the finally obtained naming rules are as follows (only identifier tokens are preserved): 
\begin{equation}
    <placeholder> \quad =\quad fetch \quad+\quad <placeholder>;  
\end{equation}
\begin{equation}
    <placeholder> \quad =\quad generate \quad+\quad <placeholder>; 
\end{equation}
\begin{equation}
    <placeholder> \quad =\quad <placeholder>s \quad+\quad next; 
\end{equation}
\begin{equation}
    <placeholder> \quad =\quad <placeholder>es \quad+\quad next;
\end{equation}
where <placeholder> denotes the tokens of variable names, and "+" concatenates the two frequently co-occurring items.
The application conditions of the first two rules are two-fold: First, the initialization should be a method call expression; Second, the method name should start with "fetch" or "generate". If these two conditions are satisfied, the variable name is likely the nouns that follow the starting verbs. For example, \emph{VarNamer} will recommend \emph{"urls"} for the initialization \emph{"generateUrls(String names)"}, and \emph{"executionStatus"} for the initialization \emph{"fetchExecutionStatus()"}.
There are three application conditions of the last two rules: First, the initialization should be a method call expression; Second, the method name should be "next"; Last, the receiver should be a plural name. If these three conditions are satisfied, the variable name is likely the singular form of the receiver. For example, \emph{VarNamer} will recommend \emph{"feature"} for the initialization \emph{"features.next()"}, and \emph{"alias"} for the initialization \emph{"aliases.next()"}.

It is worth noting that the mined rules have been submitted as pull requests~\cite{PR3,PR4} to the \emph{Eclipse} community, and the pull requests have been merged.

\subsection{Selection and Recommendation}\label{subsec:Interaction}
The names recommended by the name reuse component ($ReusedNames$) and name generation component ($GeneratedNames$) are collected, resulting in a name list, $NameList$. The name selection component works in this way:
\begin{itemize}
    \item We first validated names in $NameList$ to identify invalid entries. We considered names invalid if they fell into either of the following categories:
    \begin{itemize}
        \item Java keywords, e.g., class, for, if, static, int.
        \item Names already used in the enclosing block, e.g., local variable names or parameter names of methods.
    \end{itemize}
     Any names falling into these categories are removed from $NameList$.
    \item Following validation, if $NameList$ contains only one name, either $ReusedName$ or $GeneratedName$ is recommended. In the case where $NameList$ contains two names, i.e., a $ReusedName$ and a $GeneratedName$, the $ReusedName$ is prioritized due to the higher precision associated with the name reuse component (refer to Section~\ref{subsec:RQ4}).
\end{itemize}

Using the code snippets presented in Listing~\ref{code}, let's walk through how VarNamer recommends a name for the extract local variable refactoring:
\begin{itemize}
    \item \emph{Homogeneous Variable Retrieval:} The reuse component retrieves a homogeneous variable named \emph{"dotIdx"}. Through the filter and validator, this homogeneous variable is considered reliable for reuse.
    \item \emph{Name Generation:} Simultaneously, the generation component generates a name based on the initialization. In this case, it generates the name \emph{"lastIndexOf"}.
    \item \emph{Name Validation:} Both \emph{"dotIdx"} and \emph{"lastIndexOf"} are validated to ensure they are valid names. This step verifies whether the names are not Java keywords and are not already used within the enclosing block.
    \item \emph{Final Recommendation:} After validation, \emph{"dotIdx"} is selected from $ReusedNames$ and $GeneratedNames$ as the final recommendation. This decision is based on the fact that \emph{VarNamer} prioritizes reused names over generated ones.
\end{itemize}
By following these steps, \emph{VarNamer} ensures that the recommended name for the extracted local variable is both valid and based on established conventions or contextual information.

\section{Evaluation}\label{sec:Evaluation}
\subsection{Research Questions}
\begin{itemize}
    \item \textbf{RQ3:} How does \emph{VarNamer} perform in recommending names for extracted variables compared with baselines?
    \item \textbf{RQ4:} How do the major components contribute to the performance of \emph{VarNamer}?
    \item \textbf{RQ5:} How does \emph{VarNamer} perform regarding time efficiency?
    \item \textbf{RQ6:} How does \emph{VarNamer} work on  programming languages other than Java?
    \item \textbf{RQ7:} To what extent can \emph{VarNamer} aid developers in conducting extract local variable refactoring?
\end{itemize}

RQ3 focuses on evaluating the effectiveness of \emph{VarNamer} compared to selected baselines in recommending names for the extract local variable refactoring tasks. Additionally, we aim to uncover the underlying factors contributing to \emph{VarNamer}'s superior performance over the baselines. By addressing RQ3, we gain insights into how \emph{VarNamer} performs in real-world scenarios and the specific areas where it excels compared to baseline approaches. Understanding the advanced improvements of \emph{VarNamer} can inform future enhancements and optimizations to enhance its efficacy further.

RQ4 aims to determine the extent to which the major components of \emph{VarNamer}, namely name reuse, name generation, and name selection, contribute to its overall performance. By dissecting the roles and effectiveness of these components, we gain insights into the inner workings of \emph{VarNamer} and its capacity to recommend satisfactory variable names. Understanding the individual contributions of these components is crucial for optimizing \emph{VarNamer} and refining its functionality to better serve developers' needs in code refactoring tasks.

RQ5 focuses on evaluating the time efficiency of \emph{VarNamer} in comparison to baseline approaches, aiming to determine whether \emph{VarNamer} can meet the in-time needs of integrated development environments (IDEs) and developers. In addressing RQ5, we employ specific metrics, i.e., the time taken to recommend variable names, to gauge the time efficiency of \emph{VarNamer} and the baselines. By comparing the performance of \emph{VarNamer} against baseline approaches, we aim to elucidate its ability to deliver timely and efficient variable name recommendations, thereby supporting seamless and productive software development workflows. 

RQ6 delves into the applicability of \emph{VarNamer} beyond the Java programming language and aims to determine whether its performance remains consistent across different programming languages. In this research question, we evaluate \emph{VarNamer} in languages beyond Java, e.g., C++. By investigating RQ6, we may reveal its potential as a language-agnostic tool for enhancing developer productivity.

Having examined the performance, efficacy, and extensibility of \emph{VarNamer} in recommending variable names through preceding research questions, we now turn our focus to RQ7. While accurate recommendations are essential, their true value lies in their practical utility for developers. We consider two metrics to gauge the effectiveness of \emph{VarNamer} in assisting developers. These metrics encompass factors such as time savings during refactoring tasks, and the reduction in manual effort required for variable naming. By investigating RQ7, we may reveal the real-world impact of \emph{VarNamer} on developers' productivity.

\subsection{Dataset}~\label{subsec:overallDataset}
To evaluate the performance of \emph{VarNamer} and the baselines in recommending names for the extract local variable refactorings, we constructed a real-world refactoring dataset, called $TestingDataSet$, as specified in Section~\ref{subsubsec:DataCollection}. The dataset contains 27,158 real-world extract local variable refactorings that were mined automatically by RefactoringMiner~\cite{RefactoringMiner} from the top 1,000 Java open-source projects (sorted by stars) in GitHub. This selection of projects was made to ensure a diverse and representative sample of real-world refactorings. Notably, to guarantee the high quality of the resulting dataset, we designed a series of rules to filter out the false positives of RefactoringMiner. We would investigate research questions RQ3-5 with this dataset.

To investigate RQ6, i.e., how the proposed approach works on programming languages other than Java, we should build another dataset by collecting real-world extract local variable refactorings from source applications in other programming languages, e.g., C++.  However, to the best of our knowledge, there are no automatic tools that could discover the extract local variable refactorings in C++ applications. To this end, we manually constructed a dataset for C++ (noted as $C++Dataset$), which also serves as the dataset for investigating RQ7.  The process of the data construction is as follows:
\begin{itemize}
    \item First, we selected five well-known open-source projects from Google and Apache (i.e., "arrow"~\cite{arrow} and "mesos"~\cite{mesos} from Apache, and "angle"~\cite{angle}, "dawn"~\cite{dawn}, and "skia"~\cite{skia} from Google) on GitHub. These five projects are all active and well-maintained (with over 15,000 commits). In addition, they are across different domains including graphics engine, webGPU implementation, cluster manager, and development platform. 
    \item Second, we tracked the commit history of the selected projects and analyzed the hunks of each commit manually to collect the extract local variable refactorings.
    \item Note that the manual analysis is time-consuming and labor-intensive. To this end, for each project, we collected 10 extract local variable refactorings in the order of their appearance, resulting in 50 real-world refactorings in total. Note that there may be multiple extract local variable refactorings targeting the same expression in a single commit. To improve the quality of the dataset, we only kept two instances of them.
\end{itemize}

As a conclusion, we have constructed two datasets for the evaluation, i.e., $TestingDataSet$ containing 27,158 extract local variable refactoring in Java applications, and $C++Dataset$ containing 50 extract local variable refactoring in C++ applications.

\subsection{Baseline Approaches}~\label{subsec:BaseLines}
We included the IDEs and large language model (LLM) approaches as our baselines. For IDE baselines, we selected \emph{Eclipse}~\cite{Eclipse} and \emph{IDEA}~\cite{IDEA} due to their widespread usage in the software development community and their comprehensive features for code editing and refactoring. We implemented the extract local variable refactoring function of \emph{Eclipse} and \emph{IDEA} by leveraging their plugin development framework where the internal name recommendation interface is available. Since \emph{IDEA} provides a list of names for developers, we selected the first name from the list as the final choice to ensure consistency and fairness in the comparison process.  
For LLM approaches, we selected \emph{Incoder}~\cite{Incoder}, which represents one of the state-of-the-art approaches in code completion,  particularly focusing on cloze-style inference techniques. Note that although \emph{Incoder} was originally evaluated in Python programs, it supports variable name prediction tasks in multiple programming languages, and its performance on the cloze task in Java language is even better than that in Python language as reported by Fried et al.~\cite{Incoder}. Consequently, it is suitable to take \emph{Incoder} as our baseline on variable name recommendation in Java language.  We utilized the Python implementation of the \emph{Incoder-6.7B} model obtained on Hugging Face~\cite{HuggingFace}, a widely recognized platform for accessing and sharing pre-trained models and libraries for natural language processing tasks. To evaluate the performance of \emph{Incoder}, we prepared the data fed into \emph{Incoder} as follows:
\begin{itemize}
    \item First, for each refactoring in $TestingDataSet$, we obtained the Java file where the extract local variable refactoring happened (after the refactoring).
    \item Second, we further located the methods enclosing the refactoring and replaced the names of the newly introduced variables and their references with \emph{"<infill>"} tokens, as required by \emph{Incoder}. This preprocessing step ensures that the code snippets are formatted correctly for input into \emph{Incoder}, facilitating accurate variable name prediction.
    \item Third, we input the preprocessed code snippets, containing cloze-style placeholders, into \emph{Incoder}. We extracted the generated name for the first \emph{"<infill>"} token, which corresponds to the variable declaration, from the output. For instance, in the code snippet \emph{"final int <infill> = name().lastIndexOf('.');"}, the infilled name generated by \emph{Incoder} for the placeholder was considered as the final name for comparison. This also maintains consistency in the evaluation methodology.  
\end{itemize}

\subsection{Performance Metrics}~\label{subsec:Metrics}
To measure the performance of the evaluated approaches, we adopted a series of metrics:
\begin{itemize}
    \item  \emph{\#Total Cases:}  the number of cases involved in the evaluation, i.e., the frequency of invocation of the evaluated approaches.
    \item  \emph{\#Recommendation:}  the number of cases where the evaluated approaches make a recommendation for the developers. It is worth noting that when \emph{VarNamer} fails to generate a reasonable name (i.e., both name reuse and name generation components failed to suggest any variable name), it opts to make no recommendation rather than offer a low-quality one. That is to say, \emph{VarNamer} will not make recommendations to all the data (\#total cases) as the other three baselines. Therefore, for \emph{VarNamer}, the number of recommendations (\#recommendation) may not always match the total number of cases (\#total cases). This discrepancy should be considered when interpreting the results of the evaluation.
    \item  \emph{\#Exact Match:}  the number of cases where the recommended variable names are identical to the ground truth (i.e., the expected names). Notably, achieving an exact match is crucial because in such cases, developers can readily accept the recommendations without the need for additional edits. This not only streamlines the coding process but also ensures consistency and conciseness in the code base~\cite{Deissenboeck2006}. \emph{\#Exact match} should ideally be equal to or less than \emph{\#recommendation} as an evaluated approach can only achieve an exact match if it makes a recommendation for the given case. 
\item  \emph{$EM_{Precision}$:}  the number of exact matches divided by the number of recommendations, i.e., \begin{equation}
     \label{eqt:Precision}
         EM_{Precision}= \frac{\#Exact\ Match}{\#Recommendation}
     \end{equation}
    $EM_{Precision}$ presents how often the suggested variable names are correct. The more often the suggested names are incorrect (i.e., with lower $EM_{Precision}$), the more likely developers may abandon the suggestion approach/tool. 
\item  \emph{$EM_{Coverage}$:}  the number of exact matches divided by the number of cases involved in the evaluation, i.e.,  
     \begin{equation}
     \label{eqt:recall}
         EM_{Coverage}=  \frac{\#Exact\ Match}{\#Total\ Cases}
     \end{equation}
     $EM_{Coverage}$ presents the ratio of exact match cases generated by each suggestion approach out of the total number of cases.
\end{itemize}

\begin{table}[]
\renewcommand\arraystretch{1.2}  
        \caption{Settings of \emph{VarNamer}}
        \centering
        \renewcommand{\arraystretch}{1.2}
        \begin{tabular}{lll }
         \toprule
         & \textbf{Parameter} &\textbf{Value}   \\\hline
        & ProjectNum   & 80                       \\
        & FGSim        & 0.3                  \\
        & MinConfidence   &   0.8  \\
         & IniLength           & 30 \\
          & MinSupport        &    50  \\
                                  \bottomrule
        \end{tabular}
        \label{tab:Parameters}
        \vspace{-1em}
        \end{table}
\subsection{Setup}~\label{subsec:ExperimentalSetup}
For a fair comparison, we independently evaluated each of the involved approaches, including the proposed approach and the baselines, on the same server. The setting of the server is as follows: 
\begin{itemize}
    \item Operation system: Ubuntu 18.04.1;
    \item CPU: 32 * Intel(R) Xeon(R) CPU E5-2620 v4 @ 2.10GHz; 
    \item GPU: 4*TITAN RTX (used by the baseline approach \textit{Incoder} only); 
    \item RAM: 128 GB.
\end{itemize}

\emph{VarNamer} has several parameters that need to be tuned. We first quantified the valid value range for parameters such as $ProjectNum$ ($[0, \ 374]$), $IniLength$($[0,\  66]$), and $FGSim$($[0.0,\ 1.0]$). The maximum number of projects is all the projects included in $MiningDataSet$, i.e., 374. For $IniLength$, we determined the upper bound by considering the outliers' upper bound of character length in the $EmpiricalDataSet$, calculated as Q3+1.5*IQR=66, to ensure an appropriate parameter range. The $FGSim$ is calculated by equation~\ref{eq:totalSim} which outputs a decimal between 0 and 1. The step sizes for $ProjectNum$ and $IniLength$ were set to 10, while the step size for $FGSim$ was set to 0.1. These step sizes were chosen to facilitate a systematic exploration of parameter values. Subsequently, we employed the grid searching algorithm with the specified value ranges and step sizes to identify the parameter values that optimize the $EM_{Precision}$ and $EM_{Coverage}$ of \emph{VarNamer}. $MinSupport$ controls the number of considered frequency items. With the increase of $MinSupport$, considered frequency items decrease, leading to less useful co-appearing relationships and less time cost. $MinConfidence$ decides the reliability of the mined relationships. As a result, we empirically tuned the two parameters to make a balance between the time cost and the mined results. The finally adopted parameters are presented in Table~\ref{tab:Parameters}.

\begin{table}[]
 \renewcommand\arraystretch{1.2}
\centering
\caption{Improving the State of the Art}
\begin{tabular}{@{}lcccc@{}}
\toprule
 Metrics & Eclipse & IDEA & Incoder & VarNamer \\ \midrule
\#Recommendation   &   27,158   &  27,158   &  27,158      & 21,766    \\ 
\#Exact Match      &   7,380    &  8,022    &  9,006       & 9,036   \\ 
$EM_{Precision}$    & 27.2\%           & 29.5\%       & 33.2\%        & 41.5\%                \\ 
$EM_{Coverage}$       & 27.2\%         & 29.5\%              & 33.2\%   & 33.3\%                \\ \bottomrule
\end{tabular}
\label{tab:Performance1}
\end{table}

\begin{table}[]
 \renewcommand\arraystretch{1.2}
\centering
\caption{Performance Comparison in $CommonDataSet$}

\begin{tabular}{@{}lcccc@{}}
\toprule
 Metrics & Eclipse & IDEA & Incoder & VarNamer \\ \midrule
 \#Recommendation   &  21,766   &  21,766   &  21,766   & 21,766     \\
 \#Exact Match      & 7,052     &  7,509    &  7,640    & 9,036     \\
$EM_{Precision}$    & 32.4\%           & 34.5\%       & 35.1\%        & 41.5\%                \\ 
$EM_{Coverage}$       & 32.4\%         & 34.5\%         & 35.1\%        & 41.5\%                \\ \bottomrule
\end{tabular}

\label{tab:Performance2}
\end{table}

\subsection{RQ3: Improving the State of the Art}~\label{subsec:RQ3}
To answer RQ3, we evaluated the selected approaches on $TestingDataSet$ independently. The evaluation results are presented in Table~\ref{tab:Performance1}. 
The second to fourth columns in Table~\ref{tab:Performance1} present the performance of the selected baselines on the given testing data (i.e., $TestingDataSet$). The last column presents the performance of the proposed approach, i.e.,\emph{VarNamer}. The second row denotes the number of cases where the evaluated approaches made suggestions whereas the third row presents the number of cases where the variable names suggested by the evaluated approaches are identical to the ground truth. The fourth and the fifth rows present the $EM_{Precision}$ and $EM_{Coverage}$, respectively. 

From Table~\ref{tab:Performance1} we make the following observations:
\begin{itemize}
    \item \emph{VarNamer} may refuse to recommend in some cases while the baselines will recommend in all the cases. Although \emph{VarNamer} makes the fewest number of recommendations, i.e., 21,766 vs. 27,158, the number of its exact-match names is the highest, i.e., 9,036 vs. 7,380 for \emph{Eclipse}, 8,022 for \emph{IDEA}, and 9,006 for \emph{Incoder}.  
    \item \emph{VarNamer} outperforms all the baseline approaches by a significant margin. Compared with one of the state-of-the-art methods in recommending variable names, i.e., \emph{Incoder}, \emph{VarNamer} improves the $EM_{Precision}$ by 25.0\%=(41.5\%-33.2\%)/33.2\% without sacrificing $EM_{Coverage}$.
    \item Compared with the current implementations of popular IDEs such as \emph{Eclipse} and \emph{IDEA} for recommending variable names, \emph{VarNamer} significantly improves the $EM_{Precision}$ by 52.6\%=(41.5\%-27.2\%)/27.2\% (compared to \emph{Eclipse}) and 40.7\%=(41.5\%-29.5\%)/29.5\% (compared to \emph{IDEA}). 
\end{itemize}

Since \emph{VarNamer} works on a subset of 21,766 instances, we also report how the other baselines perform on this same subset, i.e., $CommonDataSet$ in Table~\ref{tab:Performance2}. The table structure of  Table~\ref{tab:Performance2} is the same as Table~\ref{tab:Performance1}. From  Table~\ref{tab:Performance2} we make the following observations:
\begin{itemize}
    \item When evaluated on the same sub-dataset, i.e., $CommonDataSet$, $EM_{Precision}$ and $EM_{Coverage}$ are identical for all the four approaches since the number of recommendations is equal to the number of total cases. 
    \item \emph{VarNamer} can still achieve the best $EM_{Precision}$ and $EM_{Coverage}$, i.e., 41.5\%, compared to the other three baselines, outperforming \emph{Incoder} in $EM_{Precision}$ and $EM_{Coverage}$ by 6.4 percentage points.
    \item The $EM_{Coverage}$ of \emph{VarNamer} improves from 33.3\% to 41.5\% as the number of recommendations decrease.
    \item The $EM_{Precision}$ and $EM_{Coverage}$ of \emph{Eclipse} and \emph{IDEA} improve by 5.2 percentage points and 5.0 percentage points, respectively while \emph{Incoder} only improves by 1.9 percentage points. 
\end{itemize}

To investigate the reason why \emph{VarNamer} outperforms the baseline approaches, we conducted a data analysis, and the results suggest that there are 2,260 and 2,267 cases where \emph{VarNamer} succeeded in recommending a satisfying name while \emph{Eclipse} and \emph{IDEA} failed, respectively.  
81.3\%(=1,838/2,260) of the names recommended by \emph{Eclipse} is deemed sub-optimal due to the absence of the reuse component present in \emph{VarNamer}. The remaining 18.7\%(=422/2,260) of the recommended names are sub-optimal because of the missing generation rules. For \emph{IDEA}, 77.5\% (=1,757/2,267) of the recommended names are sub-optimal due to the absence of the reuse component, while 22.5\%(=510/2,267) are sub-optimal due to missing generation rules.
We illustrate the advancement of the reuse component with listing~\ref{code} in Section~\ref{sec:Background}. In this example, \emph{VarNamer} retrieved a reliable homogeneous variable from the sibling method in the same Java file and reused its name. Consequently, \emph{VarNamer} recommended \emph{"dotIdx"}, which is identical to the name given by the original developers. However, \emph{Eclipse} recommended \emph{"lastIndexOf"} and \emph{IDEA} recommended \emph{"x"}, which are both sub-optimal. 
The advancements of the generation component in VarNamer encompass two key aspects:
First, to recommend names for method calls, both \emph{Eclipse} and \emph{IDEA} take the heuristic rule which removes the popularly starting verbs of the method names and only keeps the trailing content. To ensure time efficiency, \emph{Eclipse}, and \emph{IDEA} employ a heuristic approach to retain a predefined set of popularly used verbs, such as \emph{{"get", "is", "to"}} for \emph{Eclipse} and \emph{{"get", "is", "to", "find", "create", "as"}} for \emph{IDEA}, respectively. However, through the mining process detailed in Section~\ref{subsubsec:MiningNamingPatterns}, we discovered a significant number of popular verbs commonly used in programming, such as \emph{"read", "build", "add", "parse"}. This expanded understanding of common programming verbs enhances \emph{VarNamer}'s ability to recommend variable names that are preferred by developers.
For example, for \emph{initialization} \emph{"buildId(serviceInstance)"}~\cite{RefactoringExample2}, the involved method declaration is presented in Listing~\ref{code1}. 
\begin{lstlisting}[float,style=JavaStyle,caption={Extract Local Variable Refactoring from ConsulServiceDiscovery.java},label={code1}]
public void unregister(ServiceInstance serviceInstance) throws RuntimeException {
-   client.agentServiceDeregister(`\textbf{buildId(serviceInstance)}`);
+   String `\textbf{id}` = `\textbf{buildId(serviceInstance)}`;
    ...
+   client.agentServiceDeregister(`\textbf{id}`);
}
private String buildId(ServiceInstance serviceInstance) {
    // let's simply use url's hashcode to generate unique service id for now
    return Integer.toHexString(serviceInstance.hashCode());
}
\end{lstlisting}
As we can see from this code snippet, the functionality of the method \emph{"buildId"} is to generate a unique ID for a service object and return it. As a result, developers prefer the variable name 'id' in this context. For this case, \emph{VarNamer} removed the verb \emph{"build"} and recommended \emph{"id"} as the variable name, aligning better with the developer preferences.
However, \emph{Eclipse} recommended \emph{"buildId"} and \emph{IDEA} recommended \emph{"s"}, both of which are sub-optimal.
Additionally, \emph{VarNamer} excels in cases where initializations involve accessing elements from a collection. For example, for initialization \emph{"features.next()"}~\cite{RefactoringExample3}, the involved method declaration is presented in Listing~\ref{code3}. As we can see from this code snippet, \emph{"next"} is a method of \emph{"FeatureIterator"}, and it returns an element from a collection. Consequently, \emph{"feature"}, which is the singular form of \emph{"features"}, is preferred by developers.  With an exploration of real-world refactoring data, \emph{VarNamer} correctly identified \emph{"feature"} as the preferred variable name. However, \emph{Eclipse} and \emph{IDEA} both recommended \emph{"next"} as the variable name, which is sub-optimal.
\begin{lstlisting}[float,style=JavaStyle,caption={Extract Local Variable Refactoring from TemplatedItemsConverter.java},label={code3}]
public void writeInternal(AbstractItemsResponse itemsResponse, HttpOutputMessage httpOutputMessage) throws IOException {
    try (FeatureIterator features = itemsResponse.getItems().features()) {
        while (features.hasNext()) {
-           builder.evaluate(writer, new TemplateBuilderContext(`\textbf{features.next()}`));
+           // lookup the builder, might be specific to the parent collection
+           Feature `\textbf{feature}` = `\textbf{features.next()}`;
            ...
+           builder.evaluate(writer, new TemplateBuilderContext(`\textbf{feature}`));
        }
    }
}
\end{lstlisting}

To investigate why \emph{VarNamer} outperforms \emph{Incoder}, we randomly sampled 356 methods from 4,767 cases where \emph{VarNamer} succeeded in recommending a satisfying name while \emph{Incoder} failed, with a confidence level of 95\% and a margin of error of 5\%.
Two authors independently inspected the sampled cases and tried to find out any possible reasons that could explain why sometimes \emph{Incoder} failed and \emph{VarNamer} succeeded in recommending a correct name. The Cohen's kappa coefficient of agreement between the two authors is 0.82. Any discrepancies were resolved through discussion until a consensus was reached.
Compared to \emph{VarNamer}, \emph{Incoder} may not thoroughly leverage the \emph{initialization} context, which is essential in recommending a name for it. For example, on 299 out of the 356 samples, the extracted expressions are method invocations, and thus method invocations were used as initialization for the newly introduced variables. In such cases, our approach successfully suggested the expected variable name by extracting tokens from method names in the extracted method invocation. However, \emph{Incoder} failed in a majority of cases (161 out of 299). We also noticed that such cases (161) accounted for nearly half of the analyzed samples (356 in total), suggesting that they pose significant challenges for \emph{Incoder}. For example, for initialization \emph{"checkConfig.getMessages()"}~\cite{RefactoringExample4}, the involved method declaration is presented in Listing~\ref{code2}.  As we can see from the code snippet, The functionality of the method \emph{"getMessages"} is to return a HashMap Object containing custom messages. Consequently, \emph{"messages"} here is preferred by developers. However, \emph{Incoder} recommended \emph{"project"} even though \emph{"messages"} can be simply extracted from the method name. This challenge may be solved by fine-tuning \emph{Incoder} with our data. However, since \emph{Incoder} is trained on high-performance GPU devices (such as dual GPUs, which are not typically available to most developers), its performance in real-world scenarios would likely be even worse than what is reported. Consequently, we did not fine-tune \emph{Incoder} in this paper. It is also worth noting that \emph{Incoder} may be inflated due to data leakage. To figure out the situation of data leakage, we manually checked the involved projects, and the results suggest that 77.3\% (=576/745) of the projects used in this paper appeared in the training data of \emph{Incoder}. Despite the data leakage, \emph{VarNamer} still outperforms \emph{Incoder}, demonstrating its effectiveness.  

\begin{lstlisting}[float,style=JavaStyle,caption={Extract Local Variable Refactoring from TypeNameTest.java},label={code2}]
public void testTypeName() throws Exception {
    ...
+   final Map<String, String> `\textbf{messages}` = `\textbf{checkConfig.getMessages()}`;
-   "71:12: " + getCheckMessage(`\textbf{checkConfig.getMessages()}`, msgKey, "Annotation$", format),
+   "71:12: " + getCheckMessage(`\textbf{messages}`, msgKey, "Annotation$", format),
    ...
}
/** The map containing custom messages. */
private final Map<String, String> messages = new HashMap<>();
...
/* @return unmodifiable map containing custom messages */
@Override
public Map<String, String> getMessages() {
    return new HashMap<>(messages);
}
\end{lstlisting}
\emph{VarNamer}'s failure to recommend satisfying names can be attributed to the imperfect $EM_{Precision}$ in both the reuse and generation components (80.3\% for the reuse component and 34.9\% for the generation component).  Furthermore, \emph{VarNamer} cannot incorporate developers' naming preferences, such as the length (prefer shorter or longer names) and format (prefer full names or abbreviations) of variable names. Integrating these preferences is essential for ensuring that recommended names align with developers' expectations and coding conventions.

\begin{table}[]
\renewcommand\arraystretch{1.2}  
    \centering
    \caption{Contributions of VarNamer's Major Components}
    \begin{tabular}{@{}ccc@{}}
    \toprule
             & $EM_{Precision}$ & $EM_{Coverage}$\\ \midrule
    VarNamer &  41.5\% &  \textbf{33.3\%}  \\
    w/o Name Generation  & \textbf{80.3\%} &   10.7\%    \\
    w/o Name Generation [Additional Rules] & 40.8\% & 30.2\% \\
    w/o Name Reuse  &  34.9\% &   27.2\%   \\
    w/o Name Selection  &  35.3\% &   28.8\%   \\\bottomrule
    \end{tabular}
    \label{tab:AblationStudy}
    \vspace{-1em}
    \end{table}

\begin{tcolorbox}
        
        \textit{Answer to RQ3:}
        \emph{VarNamer} significantly improves the $EM_{Precision}$ by 52.6\% (compared to \emph{Eclipse}) and 40.7\% (compared to \emph{IDEA}).  Compared with \emph{Incoder}, \emph{VarNamer}  improves the $EM_{Precision}$ by 25\% without sacrificing $EM_{Coverage}$.
        
    \end{tcolorbox}

\subsection{RQ4: Contributions of Major Components of VarNamer}~\label{subsec:RQ4}
To address this research question, we removed each component successively, resulting in three variants of \emph{VarNamer}. We then evaluated each variant's performance on the $TestingDataSet$ and compared it to the baseline performance of VarNamer with all components enabled.
To investigate the influence of the name selection component, we removed the name validation and adopted a random selection strategy to select the final name for recommendation. To figure out the real contributions of the additional rules proposed in this paper, we also excluded the additional rules mined from a large code corpus to avoid potential overlap with the existing rules adopted by \emph{Eclipse} and \emph{IDEA}.

The evaluation results presented in Table~\ref{tab:AblationStudy} reveal the following observations:
\begin{itemize}
    \item All three major components make significant contributions to the performance of \emph{VarNamer}, as evidenced by the significant decrease in performance when any of them is removed.
    \item The name reuse component contributes more to the $EM_{Precision}$, with a decrease of 15.9\%=(41.5\%-34.9\%)/41.5\% in $EM_{Precision}$ when it is removed.  The $EM_{Precision}$ of \emph{VarNamer} remains high at 80.3\% even when the name generation component is removed, providing additional support for the significant contribution of the name reuse component.
    \item The name generation component contributes more to the $EM_{Coverage}$, with a decrease of 67.9\%=(33.3\%-10.7\%)/33.3\% when it is removed. The additional rules in this component also play a crucial role, as demonstrated by a decrease of 9.3\%=(33.3\%-30.2\%)/33.3\% in $EM_{Coverage}$ when these rules are removed.
    \item The name selection component is of great importance to \emph{VarNamer}. when it is removed, the $EM_{Precision}$ decreases by 14.9\%=(41.5\%-35.3\%)/41.5\% and the $EM_{Coverage}$ decreases by 13.5\%=(33.3\%-28.8\%)/33.3\%.
\end{itemize}

It is worth noting that we did not incorporate data type in our approach although data type is one of the most critical contexts where name tokens of variables can be found. We investigated the impact of data type on the \emph{VarNamer}'s performance and the results are presented in Table~\ref{tab:DataTypeImpact}. Based on the current implementation of \emph{VarNamer}, we further developed two variants, i.e., "w/ Data Type" and "w/ Data Type + Selection Strategy". These two variants indicate incorporating data types without any selection strategy and with a selection strategy, respectively. The selection strategy involves making recommendations only for cases where the types are customized by developers, as our observations suggest that developers are highly likely to use user-defined types directly as variable names.

From Table~\ref{tab:DataTypeImpact}, we make the following observations:
\begin{itemize}
    \item Incorporating data type into \emph{VarNamer} without any selection strategy improves $EM_{Coverage}$ by 1.6 percentage points, while $EM_{Precision}$ significantly drops by 5.9 percentage points. 
    \item With the selection strategy, $EM_{Coverage}$ still improves by 1.5 percentage points, while the $EM_{Precision}$ drops by 1.2 percentage points. A trade-off exists between  $EM_{Coverage}$ and $EM_{Precision}$ when incorporating data type into \emph{VarNamer}.
    \item Ideally, adding more effective selection strategies could further improve the $EM_{Coverage}$ without sacrificing too much $EM_{Precision}$. However, in this paper, we choose to not incorporate data type because we prioritize $EM_{Precision}$.
\end{itemize}

The possible reason that adding data types reduces $EM_{Precision}$ is due to their high overlap with other types of contexts, even though the chance of an exact match between data types and the expected variable names is 13.8\% as shown in Table~\ref{tab:TotalSame}. Our investigation found that 80.9\% of sub-tokens found in \emph{data types} also appear in \emph{initializations}. Additionally, the overlapping with \emph{homogeneous variables} and \emph{assignments} are 16.5\% and 0.9\%. Overall, only 6.2\% sub-tokens can be found exclusively in \emph{data types}.     

\begin{table}[]
\renewcommand\arraystretch{1.2}  
    \centering
    \caption{Impact of data type on the VarNamer's Performance}
    \begin{tabular}{@{}ccc@{}}
    \toprule
         Setting    & $EM_{Precision}$ & $EM_{Coverage}$\\ \midrule
    VarNamer &  41.5\% &  33.3\%  \\
    w/ Data Type  & 35.6\% &   34.9\%    \\
    w/ Data Type + Selection Strategy  & 40.3\% &   34.8\%     \\\bottomrule
    \end{tabular}
    \label{tab:DataTypeImpact}
    \vspace{-1em}
    \end{table}

\begin{tcolorbox}
        
        \textit{Answer to RQ4:}
        All three major components make significant contributions to the performance of \emph{VarNamer}. $EM_{Precision}$ benefits more from the name reuse component whereas $EM_{Coverage}$ benefits more from  the name generation component. Incorporating data types improves the $EM_{Coverage}$ with a sacrifice of $EM_{Precision}$.
        
    \end{tcolorbox}

\subsection{RQ5: Time Efficiency}~\label{subsec:RQ5}

In this research question, we investigated the time cost of \emph{VarNamer} and the three baselines on $TestingDataSet$, and the comparison results are presented in Table~\ref{tab:TimeCost}. The second column denotes the total running time cost. The third column denotes the average running time cost for a single refactoring, and the last column denotes the median value of the running time cost for a single refactoring. It's important to note our methodology: we ran all the approaches (except \emph{Incoder}) five times repeatedly and took an average of the total time, the average time per refactoring, and the median time per refactoring. Since the name recommendation of \emph{Incoder} is time-consuming, and there is nothing comparable between the time cost of \emph{Incoder} and that of the other three approaches, we only executed \emph{Incoder} once and recorded its time cost for one run.
From Table~\ref{tab:TimeCost}, we make the following observations:
\begin{itemize}
    \item Considering the average time cost, \emph{Incoder}~\cite{Incoder} takes the most time (48.4s) to generate a name for an extracted local variable, while \emph{VarNamer}, \emph{Eclipse}~\cite{Eclipse}, and \emph{IDEA}~\cite{IDEA} take much shorter time (5.7ms, 0.5ms, and 3.9ms). Concerning the median value, the time cost is 27.0s vs. 5.0ms, 0.2ms, and 3.0ms. This is reasonable because \emph{Incoder} depends on deep neural networks that possess extremely complex structures and billions of parameters (6.7 billion), while the other three approaches utilize efficient heuristic rules based on lexical and syntactic patterns.
    \item It takes \emph{VarNamer} an average of 5.7ms to recommend a name, which is comparable to that of \emph{IDEA} (3.9ms). This indicates that \emph{VarNamer} is efficient enough to be incorporated into popular IDEs.
\end{itemize}

In conclusion, we underscore the importance of considering both efficiency and accuracy when evaluating name recommendation approaches. While deep learning-based methods like \emph{Incoder} may offer high accuracy, their efficiency may be compromised. Conversely, the proposed approach \emph{VarNamer} strikes a balance between efficiency and accuracy, making it a suitable candidate for integration into mainstream IDEs.

\begin{table}[]
\renewcommand\arraystretch{1.2}  
    \centering
    \caption{Efficiency of Evaluated Approaches}
        \begin{threeparttable}
    \begin{tabular}{@{}cccc@{}}
    \toprule
       Approaches      & Execution Time &  \makecell[c]{Execution Time \\per Refactoring \\(Average)} &  \makecell[c]{Execution Time \\per Refactoring \\(Median)}\\ \midrule
    Eclipse  & 11.5s                & 0.5ms       & 0.2ms              \\
    IDEA        & 84.7s                & 3.9ms       & 3.0ms           \\
    Incoder  & 292.6h               & 48.4s       & 27.0s       \\ \midrule
    VarNamer                & 123.5s               & 5.7ms       & 5.0ms \\ \bottomrule
    \end{tabular}
        \begin{tablenotes}    
                    
        \item \textit{ms}: milliseconds; \textit{s}: seconds;
         \textit{h}: hours.
  \end{tablenotes}  
        \end{threeparttable} 
    \label{tab:TimeCost}
     \end{table}

\begin{tcolorbox}
        
        \textit{Answer to RQ5:}
        \emph{VarNamer} is efficient. It can make a recommendation in 5ms,  comparable to \emph{IDEA}.  
        
    \end{tcolorbox}
    
\subsection{RQ6: Application to C++}~\label{subsec:RQ6}

The experiment results presented in Section~\ref{subsec:RQ3} have indicated that \emph{VarNamer} performs well in recommending variable names in Java programming languages. However, it remains unknown whether \emph{VarNamer} can also be applied to recommend variable names in other programming languages besides Java. To investigate the extensibility of \emph{VarNamer}, we evaluated its performance in C++ programming languages in this research question. 

Theoretically, \emph{VarNamer} is language-agnostic because the factors (i.e., homogeneous variables, initializations of variables) are common to most programming languages. However, to the best of our knowledge, there is not a universal AST parser for all programming languages. Due to the absence of a universal AST parser, we had to implement a C++ version of \emph{VarNamer} using an AST parser specifically designed for C++. Our approach was developed based on the \emph{Eclipse CDT}~\cite{cdt}, which provides a fully functional C and C++ Integrated Development Environment on the \emph{Eclipse} platform. By incorporating the core logic of our approach into this framework, we obtained the C++ version of our approach and called \emph{VarNamer-C++} for convenience. We evaluate \emph{VarNamer-C++} on $C++Dataset$, and the process of the dataset construction is presented in Section~\ref{subsec:overallDataset}.

We leveraged the same metrics, i.e., $EM_{Precision}$ and $EM_{Coverage}$ (refer Section~\ref{subsec:Metrics}) to evaluate how \emph{VarNamer-C++} performed in recommending variable names on $C++Dataset$.
The evaluation results suggest that the $EM_{Coverage}$ and $EM_{Coverage}$ of \emph{VarNamer-C++} in recommending variable names are both 44.0\%(=22/50), which is comparable to the performance (41.5\%) in Java programming language. However, the $EM_{Precision}$ and $EM_{Coverage}$ of \emph{Eclipse CDT} is only 12.5\%(=3/24) and 6.0\%(=3/50). 

To investigate the reason why \emph{VarNamer-C++} outperforms \emph{Eclipse CDT}, we analyzed the names recommended by them in 50 refactorings. The analysis results suggest that there are two major reasons. One reason is that \emph{Eclipse CDT} declined to recommend a name in most cases (26 out of 50). We found that the recommendation logic of \emph{Eclipse CDT} for the extract local variable refactoring is quite simple and only covers a few specific cases, leading to a low ratio of recommendations. Another reason also exists in the current implementations of IDEs such as \emph{Eclipse} and \emph{IDEA}. As introduced in Section~\ref{sec:Introduction}, these IDEs ignore many useful contexts, e.g., homogeneous variables, and lack in-depth analysis of real-world refactoring data. Through analysis of real-world refactoring data, \emph{VarNamer-C++} is better equipped to cover more cases and provide recommendations that are more satisfying to developers.

Consequently, we conclude that our approach can be extensively applied to other programming languages besides Java, e.g., C++, and its performance remains stable as indicated by the experiment results.

By reusing the parameter values from the Java version of our approach, as shown in Table~\ref{tab:Parameters}, we still achieved comparable performance in recommending names for the extract local variable refactorings in C++. However, in our investigation of the differences between Java and C++ programming languages, we observed several distinctions between Java and C++ that may have a slight impact on name recommendation. For instance, in addition to the Java-style method invocation using '.', C++ offers two alternative methods: the pointer operator '->' and the scope operator '::'. Notably, the latter two operators consist of one more letter than a single '.'. Moreover, the concept of \emph{"namespace"} is crucial in C++ programming, leading to a prevalent use of scope operators like \emph{"gl::FromGLenumgl::ShadingRate(rate)"}. Developers in C++ often prefix object references with their corresponding namespace. These differences result in longer initialization (in character length) in C++ compared to Java.

\begin{tcolorbox}
        
        \textit{Answer to RQ6:}
        The performance of \emph{VarNamer}  on C++ source code is comparable to that on Java code.
        
    \end{tcolorbox}

\subsection{RQ7: User Study}~\label{subsec:RQ7}

To investigate to what extent \emph{VarNamer-C++} can enhance the efficiency of extract local variable refactoring in the wild, we conducted a user study with real-world extract local variable refactorings (i.e., $C++Dataset$) and \emph{VarNamer-C++}. Characteristics of the refactoring dataset and \emph{VarNamer-C++} are presented in Section~\ref{subsec:overallDataset} and Section~\ref{subsec:RQ6}, respectively.

We invited six developers who had development experience in C++ projects to conduct the user study. The participants were divided into two groups with three developers in each group. Both groups were asked to conduct extract local variable refactorings on $C++Dataset$, and the difference is that the first group finished it with the help of the standard CDT plugin (latest version by the time we submitted this paper, i.e., \emph{cdt-11.5.0}) and the second group finished it with the help of \emph{VarNamer-C++}. The procedure for developers conducting a single refactoring is as follows: 
\begin{itemize}
    \item For each refactoring instance in $C++Dataset$, participants were asked to first call the name recommendation dialog through shortcut keys \emph{"Alt + Shift + L"} in \emph{Eclipse}. 
    \item Then the participants were asked to judge whether the recommended name was proper in the given context (i.e., the enclosing method declarations). 
    \item If so, they clicked \emph{"OK"} and finished a single refactoring. Otherwise, they were asked to edit the recommended name until they were satisfied with it. In addition, if the dialog did not contain a name, i.e., no recommendation is available, they were asked to coin a name from scratch for the newly introduced variable. After giving a name to the newly introduced variable, the participants clicked \emph{"OK"} and finished a single refactoring.
    \item We recorded the time of each developer finishing all the refactorings and the names they finally selected for the newly introduced variables.
\end{itemize}

We leveraged two metrics, i.e., time cost and edit distance, to measure the ability of \emph{VarNamer-C++} to improve the development efficiency. Time cost represents the time developers take to finish all the refactorings. The edit distance (between the names recommended and the names finally selected by developers) reflects developers' satisfaction with the recommended names. These operations encompass replacing, inserting, and deleting a letter, providing a suitable measure of developers' editing efficiency.

To avoid the unfairness of this experiment, we took the following measures:
\begin{itemize}
    \item We ensured that the experience of developers in each group was evenly distributed (the average and median years of development experience for both groups are 2.3 and 2 years, respectively). 
    \item To minimize the interference of irrelevant factors, we completed all necessary preparations before the experiments began. This included opening \emph{Eclipse}, installing CDT and \emph{VarNamer-C++}, opening files, and locating the expressions to be extracted. In addition, all six developers conducted the refactorings on the same personal computer to avoid unnecessary interference. Additionally, developers were unfamiliar with the selected C++ files, and all project information was anonymous.
    \item Due to the heavy workload of development tasks, developers were instructed to complete each refactoring within 60 seconds to simulate real-world development scenarios.
\end{itemize}

\begin{table}[]
\renewcommand\arraystretch{1.2}  
\centering
\caption{User Study}
\begin{tabular}{@{}llccccc@{}}
\toprule
Metrics                               &   Grouping     & Participant A & Participant B & Participant C & Average & Median \\ \midrule
\multirow{2}{*}{\makecell[c]{Time Cost \\(seconds)}} & Group1 (with Eclipse CDT) & 1,822  & 1,850   & 1,534   & 1,735  & 1,822 \\ \cline{2-7}
                               & Group2 (with VarNamer-C++) & 1,431  & 1,076   & 1,253   & 1,253  & 1,253       \\ \midrule
\multirow{2}{*}{Edit Distance} & Group1 (with Eclipse CDT)& 290    &  319    & 309     & 306    & 309       \\\cline{2-7}
                               & Group2 (with VarNamer-C++)& 117    &  206    & 141     & 155    & 141      \\ \bottomrule
\end{tabular}
\label{tab:LiveStudy}
\end{table}

The results are presented in Table~\ref{tab:LiveStudy}. From this table, we make the following observations:
\begin{itemize}
    \item Developers in \emph{Group2} finished the task more quickly than those in \emph{Group1}. On average, developers in \emph{Group2} completed the refactorings 482 seconds earlier than those in \emph{Group1}. Using \emph{VarNamer-C++} speeds up extract local variable refactorings by 27.8\% compared to CDT, saving 482 seconds out of a total of 1,735 seconds.
    \item Developers in \emph{Group2} made fewer edits than those in \emph{Group1}. On average, developers in \emph{Group2} made 151 fewer edits than those in \emph{Group1}. Using \emph{VarNamer-C++} reduces edits on recommended variable names by 49.3\%, which means 151 fewer edits out of a total of 306 edits.
    \item 
    Before conducting the t-test analysis on the evaluation metrics, we first conducted Shapiro-Wilk test and Levene test to make sure that the two groups of data (1) conform to a normal distribution and (2) satisfy the homogeneity of variances. The test results are presented in Table~\ref{tab:TestAnalysis}. The results (i.e., all the p-values are greater than 0.05) suggest that the involved data satisfy the prerequisite condition of the t-test. We then conducted a t-test, which yielded a statistic of 3.35 and a p-value of 0.03 for time cost, and a statistic of 5.42 and a p-value of 0.01 for edit distance. These results indicate significant differences between \emph{Group1} and \emph{Group2} in terms of both metrics.
    
\end{itemize}

\begin{table}[]
\renewcommand\arraystretch{1.2}  
\centering
\caption{Results of Prerequisite Condition Inspection on T-test}

\begin{threeparttable}
\begin{tabular}{@{}ccccc@{}}
\toprule
Metrics                               &   Items       & Shapiro-Wilk Test & Levene Test & T-test \\ \midrule
\multirow{2}{*}{\makecell[c]{Time Cost}} & statistics & 0.99/0.82   & 0.01   & 3.35  \\ \cline{2-5}
                               & p-value  & 0.99/0.15   & 0.91   & 0.03    \\ \midrule
\multirow{2}{*}{Edit Distance} & statistics   &  0.97/0.93    & 1.03     & 5.42          \\\cline{2-5}
                               & p-value  &  0.66/0.50    & 0.37     & 0.01        \\ \bottomrule
\end{tabular}
        \begin{tablenotes}    
                  
        \item \textit{*/*}: the statistics or p-value of two groups of data for the conformity to normal distribution
  \end{tablenotes}  
        \end{threeparttable} 
\label{tab:TestAnalysis}
\end{table}

In conclusion, we underscore the substantial benefits of adopting \emph{VarNamer-C++} in the context of the extract local variable refactorings. Not only does \emph{VarNamer-C++} significantly enhance efficiency by reducing time cost, but it also improves developers' satisfaction by minimizing the need for extra edits of the recommended names. We highlight \emph{VarNamer-C++} as an effective tool for improving the productivity of developers.

\begin{tcolorbox}
        
        \textit{Answer to RQ7:}
        In terms of time efficiency, \emph{VarNamer-C++} speeds up extract local variable refactorings by 27.8\%.
        In terms of edit efficiency, \emph{VarNamer-C++} reduces edits on recommended variable names by 49.3\%.
        
    \end{tcolorbox}

\section{Discussion}\label{sec:Discussion}
\subsection{Threats to Validity}

The threat to external validity arises from potential discrepancies between RefactoringMiner's criteria for identifying extract local variable refactorings and our criteria, which could result in false positives and affect the validity of our experiments. To address this concern, we devised a set of rules to automatically filter out these false positives. These rules were carefully crafted based on the characteristics and patterns of extract local variable refactorings in our dataset. They were designed to identify and exclude instances that did not align with our specific definition of an extract local variable refactoring. By implementing these rules, we aimed to ensure the accuracy and reliability of the refactorings included in our dataset, thus enhancing the validity of our experimental results.

Another potential threat to external validity is that our evaluation of \emph{VarNamer} was limited to a large real-world refactoring dataset for Java, leaving uncertainty about its performance in recommending names for extract local variable refactorings in other programming languages. To mitigate this threat, we took proactive steps to evaluate \emph{VarNamer-C++} on a manually constructed small-scale real-world refactoring dataset specifically tailored for C++. This dataset was carefully curated to include a representative sample of C++ refactorings. We then rigorously assessed the performance of \emph{VarNamer-C++} on this dataset, ensuring that our findings could be generalized beyond the Java context. This approach provides insights into the cross-language applicability and effectiveness of our approach, enhancing the external validity of our study.

The threat to internal validity arises from the potential impact of the parameters used in our study on the performance of \emph{VarNamer}. To address this concern, we conducted parameter tuning on a separate dataset, namely $EmpiricalDataSet$, rather than on the testing dataset $TestingDataset$. This approach ensures that the performance of \emph{VarNamer} remains stable even if changes occur in the testing data.

Additionally, another threat to internal validity stems from the implementation of the IDE baselines, namely \emph{Eclipse} and \emph{IDEA}. The reported performance of these IDEs in our paper may not perfectly mirror their performance in real-world application scenarios. To mitigate this threat, we meticulously implemented \emph{Eclipse} and \emph{IDEA} by invoking internal interfaces and providing all the necessary parameters through the plugin development framework. This approach ensures that the behavior of \emph{Eclipse} and \emph{IDEA} in our experiments closely aligns with their behavior in practical usage scenarios, enhancing the internal validity of our study.

The threat to user study validity stems from the potential influence of different development experiences on the time taken to complete the refactoring task and the preference for certain variable names. To mitigate this threat, we meticulously balanced the years of development experience among participants in each group, ensuring that both the average and median years of development experience were comparable (2.3 and 2 years, respectively). Additionally, we implemented a series of measures outlined in Section~\ref{subsec:RQ7} to guarantee the fairness of the experiment. These measures included providing clear instructions, standardizing the refactoring task, and anonymizing project information to minimize biases. By carefully controlling these factors, we aimed to create a level playing field for all participants, thus enhancing the validity of our user study results. In addition, bias may exist for raters in Section~\ref{subsec:RQ3} since they know the sample is from cases where \emph{VarNamer} outperformed \emph{incoder}. As a result, they might be biased toward finding reasons for this outcome rather than analyzing it more objectively. As a mitigation measure, we had two raters conduct the analysis independently and then discuss any disagreements until a consensus was reached. 

\subsection{Limitations}

A major limitation of VarNamer is its current performance, with $EM_{Precision}$ and $EM_{Recall}$ metrics of only 41.5\% and 33.3\%, respectively. These results indicate significant room for improvement, highlighting the challenges associated with automatically suggesting new variable names accurately. An intriguing avenue for future research is to integrate developers' naming preferences into our approach. By leveraging insights into how developers typically name variables, we can potentially enhance \emph{VarNamer}'s performance and address its current limitations more effectively. Incorporating developers' naming preferences could involve analyzing patterns in existing codebases, conducting surveys or interviews with developers to understand their naming conventions, and incorporating this knowledge into the recommendation process.

A second limitation of our approach is its reliance on heuristic rules, which were designed based on the analysis of real-world data. This approach was chosen to ensure that our solution could be seamlessly incorporated into mainstream Integrated Development Environments (IDEs), where efficiency and low latency are paramount. However, while heuristic rules provide a pragmatic solution, they may not capture all nuances of variable naming across different codebases and programming paradigms. Looking ahead, there is an opportunity to leverage the power of advanced Large Language Models (LLMs) and other AI techniques to enhance the variable name recommendation process. By training AI models on vast amounts of code and incorporating contextual understanding, these models have the potential to provide more nuanced and contextually relevant variable name suggestions. 

A third limitation of our approach is its current focus on recommending names exclusively for local variables within extract local variable refactorings. While this serves the immediate need for variable name recommendations within this specific refactoring context, it does not extend to recommending names for other identifiers such as method names and class names in related refactorings like extract method and extract class. This limitation stems from the constrained scope of our current implementation, which focuses solely on extract local variable refactorings. However, there is potential for future work to broaden the scope of our approach to encompass a wider range of refactorings that involve the extraction of code entities. By developing mechanisms to recommend names for other types of identifiers, such as methods and classes, our approach could provide more comprehensive support for developers across various refactoring scenarios.

\section{Related Work}\label{sec:RelatedWork}
\subsection{Automatic Variable Renaming}
Two notable approaches, Zhang et al.\cite{Zhang2023An} and Liu et al.\cite{RefBERT}, have been developed specifically to recommend high-quality names for variable renaming tasks. Zhang et al.\cite{Zhang2023An} introduced an identifier renaming prediction and suggestion approach that operates across different granularity levels. Their method begins by predicting whether an identifier requires renaming, then utilizes commit history and naming pattern information to propose a new name. Conversely, \emph{RefBERT}, proposed by Liu et al.\cite{RefBERT}, employs contrastive learning to recommend names for variables. This approach involves two stages: length prediction and token generation.
However, it's important to note that both of these approaches rely on the original name (i.e., the name before refactoring) as one of the inputs to their models. This differs from the task of recommending a name for the extract local variable refactoring, where the objective is to suggest a suitable name for a newly created variable. Therefore, we did not include these approaches as baselines in our study.
Additionally, there are several approaches designed to recover variable names for decompiled code, such as DIRE~\cite{lacomis2019dire}, DIRECT~\cite{nitin2021direct}, and DIRTY~\cite{chen2022augmenting}. They are designed to handle the variable renaming on deterministically derived representations of stripped binaries. Although it might be possible to pose the variable extract-variable question in a form that can be input to DIRE, DIRECT, or DIRTY, this would require substantial engineering effort and doesn't offer a clear advantage over some advanced code completion techniques, e.g., \emph{Incoder}~\cite{Incoder}. Therefore, these approaches were not within the scope of our research.

\subsection{Code Completion}
In addition to the approaches specifically designed for variable renaming tasks, there are also existing code completion tools that have the potential to be applied to the name recommendation process for extract local variable refactorings. These tools, developed by both industry and academia, aim to assist developers by suggesting variables or method calls within an Integrated Development Environment (IDE).
One of the pioneering tools in this area is \emph{Prospector}, developed by Mandelin et al.~\cite{mandelin2005jungloid}. \emph{Prospector} focuses on suggesting variables or method calls within an IDE environment to enhance developer productivity.
Subsequently, a series of tools and plugins have been proposed to further facilitate code completion tasks. These include \emph{InSynth}\cite{GveroKKP2013Complete}, which provides complete code snippets based on partial input; \emph{Sniff}\cite{chatterjee2009sniff}, which offers code suggestions based on context; \emph{IntelliCode}\cite{svyatkovskiy2021fast}, a tool developed by Microsoft that utilizes machine learning to enhance code completion suggestions; JSparrow\cite{JSparrow}, which provides intelligent code recommendations and automatic code fixes; and \emph{Codota AI Autocomplete}~\cite{CodotaAI}, an AI-powered code completion tool that suggests relevant code snippets based on context.
These tools leverage various techniques, such as pattern matching, machine learning, and code analysis, to provide accurate and contextually relevant code completion suggestions to developers during software development tasks.

In the academic community, approaches for code completion can be broadly categorized into statistical language model-based and deep learning-based approaches~\cite{Mastropaolo2023}.
Statistical language model-based approaches, as demonstrated by Tu et al.\cite{Tu2014On} and Hellendoorn et al.\cite{hellendoorn2017deep}, leverage n-gram models to enhance code completion by incorporating code features.
On the other hand, deep learning-based approaches have gained popularity for automated code completion tasks~\cite{chen2024deep,kim2021code,alon2020structural}. Kim et al.\cite{kim2021code} and Alon et al.\cite{alon2020structural} proposed methods that integrate syntactic code structures to improve code completion accuracy.
Another line of research focuses on automatically renaming variables using deep learning techniques. Liu et al.\cite{liu2020multi} and Mastropaolo et al.\cite{Mastropaolo2023,Raffel2020Exploring} pre-trained deep learning language models on large code corpora and fine-tuned them for specific code completion tasks.
However, a common limitation of these approaches is their reliance on the context preceding the completion position, which may lead to sub-optimal performance in recommending names for newly extracted variables. This task differs from traditional code completion as it involves completing the code from right to left~\cite{Incoder}, presenting a challenge for existing models.
Consequently, large language models designed for infilling arbitrary code positions have emerged as promising solutions for recommending variable names in extract local variable refactorings. These models treat variable name recommendation as a cloze task and demonstrate efficient performance in this specific task.
CodeT5, introduced by Wang et al.\cite{wang2021codet5}, is a pre-trained language model with Masked Language Modeling (MLM) as the training objective, making it suitable for code completion tasks. Similarly, UniXcoder\cite{guo2022unixcoder}, a unified multi-modal pre-trained language model, excels in both code understanding and code generation tasks, including code completion. Fried et al. presented Incoder~\cite{Incoder}, a large generative code model known for its strong performance in infilling arbitrary code regions, achieving competitive results in various code infilling and editing tasks.
In addition to these established models, emerging large language models like StarCoder~\cite{Li2023}, SantaCoder~\cite{allal2023santacoder}, Code Llama~\cite{roziere2023code}, OctoCoder (OctoGeeX)\cite{muennighoff2023octopack}, and WizardCoder\cite{luo2023wizardcoder} are also gaining attention for their capabilities in code completion tasks.
As one of the state-of-the-art models in code completion, Incoder serves as a baseline in our study, representing the latest advancements in deep learning-based approaches.

\subsection{Improvement of Identifiers' Quality}
The quality of identifiers has garnered significant attention due to its profound impact on program comprehension and software maintenance~\cite{Lawrie2006,lawrie2007quantifying,butler2009relating,host2009debugging,butler2010exploring,Allamanis2015,Lin2019On,wang2025deep,liu2023automated}. As a result, various approaches have been developed to enhance the quality of code identifiers, which can be categorized into heuristic-based, statistical language model-based, deep learning-based, information retrieval-based, and generation-based methods.
Heuristic-based approaches rely on predefined rules and patterns derived from programming conventions and common coding practices to assess identifier quality and identify inconsistencies. For instance, researchers have proposed heuristic rules for identifier quality appraisal and the detection of inconsistent identifiers~\cite{caprile2000restructuring,Alsuhaibani2021,Peruma2021,Alsuhaibani2022An,Thies2010}.
Statistical language model-based approaches utilize statistical techniques and natural language processing methods to analyze code and improve identifier quality. Allamanis et al.\cite{allamanis2014learning} and Lin et al.\cite{Lin2017Investigating} utilized n-gram language models to identify low-quality identifiers by analyzing code corpus statistics.
Deep learning-based approaches leverage neural network architectures to automatically learn representations of code and identify patterns related to identifier quality. While several approaches focus on recommending names for methods~\cite{Nguyen2020,Li2021,liu2022learning,zugner2021language,wang2021lightweight,zhu2023automating}, there are also those specifically designed to address identifier quality, including variable names.
Information retrieval-based approaches, such as those proposed by Liu et al.~\cite{Liu2019}, leverage techniques from information retrieval to recommend names for methods and identify inconsistent method names.
Generation-based approaches, similar to those mentioned in DL-based approaches, automatically generate identifier names based on learned representations of code and contextual information.
In this paper, we focus on the quality of variable names introduced by extract local variable refactorings, emphasizing the importance of accurately recommending names for newly introduced variables.

\section{Conclusions and Future Work}\label{sec:Conclusion}
Software refactoring is a common practice, and mainstream Integrated Development Environments (IDEs) offer robust tool support for executing refactoring operations. However, existing tool support primarily focuses on the automated execution of predefined refactoring solutions rather than on recommending refactoring opportunities or solutions, particularly regarding the naming of variables.
The \emph{Extract Local Variable} refactoring is a prime example where IDEs often fall short in recommending appropriate variable names, despite their proficiency in automatically modifying source code. To address this limitation, we propose \emph{VarNamer}, an automated approach for recommending names for \emph{extract local variable} refactorings by leveraging their contextual information.
In this paper, we adopt program static analysis techniques such as lexical analysis, syntax analysis, control flow, and data flow analysis, along with data-mining techniques such as the FP-growth algorithm, to explore real-world refactoring data and design our approach.
To evaluate the effectiveness of our proposed approach, we constructed two datasets comprising real-world \emph{extract local variable} refactorings from open-source applications. Our evaluation on these datasets demonstrates that \emph{VarNamer} significantly outperforms state-of-the-art IDEs. Specifically, it improves the chance of exact name matching by 52.6\% compared to \emph{Eclipse} and 40.7\% compared to \emph{IntelliJ IDEA}. Additionally, a carefully designed user study indicates that our approach accelerates the refactoring process by 27.8\% and reduces the need for manual edits by 49.3\% on recommended variable names.

Notably, the key heuristic rules of our approach have been merged into \emph{Eclipse} and distributed with its releases. Specifically, we submitted to the \emph{Eclipse} community in total four pull requests that have been approved and merged. Two pull requests~\cite{PR1,PR2} implemented the heuristics to suggest new names by reuse, and another two pull requests~\cite{PR3,PR4} implemented the heuristics to generate variable names from scratch.
In the future, we plan to extend our approach to recommend names for more refactorings like extract method and extract class. It could also be interesting to investigate how to leverage the power of advanced LLMs to further improve the name recommendation.

\section*{Acknowledgments}
The authors would like to say thanks to anonymous reviewers for their insightful comments and suggestions.
This work was partially supported by the National Natural Science Foundation of China (62232003, 62172037, and 62141209), China Postdoctoral Science Foundation (No. 2023M740078), and China National Postdoctoral Program for Innovative Talents (BX20240008).

\bibliographystyle{ACM-Reference-Format}
\bibliography{reference}

\end{document}